\documentclass[longauth]{aa}  

\usepackage{graphicx}
\usepackage{pdflscape}
\usepackage{rotating}
\usepackage[breaklinks,colorlinks,citecolor=blue,linkcolor=blue,urlcolor=blue]{hyperref}
\usepackage{txfonts}
\newcommand{\HII}{H\,{\sc ii}\rm\,}
\newcommand{\NII}{{[N\,{\sc ii}]}}
\newcommand{\NIIs}{{[N\,{\sc ii}]\,}}
\newcommand{\NIIl}{{[N\,{\sc ii}]\,$\lambda$}}
\newcommand{\NeIII}{{[Ne\,{\sc iii}]}}
\newcommand{\NeIIIs}{{[Ne\,{\sc iii}]\,}}
\newcommand{\NeIIIl}{{[Ne\,{\sc iii}]\,$\lambda$}}
\newcommand{\SII}{{[S\,{\sc ii}]}}
\newcommand{\SIIs}{{[S\,{\sc ii}]\,}}
\newcommand{\SIIl}{{[S\,{\sc ii}]\,$\lambda$}}
\newcommand{\SIIll}{{[S\,{\sc ii}]\,$\lambda\lambda$}}

\newcommand{\OIII}{{[O\,{\sc iii}]}}
\newcommand{\OIIIs}{{[O\,{\sc iii}]\,}}
\newcommand{\OIIIl}{{[O\,{\sc iii}]\,$\lambda$}}
\newcommand{\OII}{{[O\,{\sc ii}]}}
\newcommand{\OIIs}{{[O\,{\sc ii}]\,}}
\newcommand{\OIIl}{{[O\,{\sc ii}]\,$\lambda$}}
\newcommand{\OIIll}{{[O\,{\sc ii}]\,$\lambda\lambda$}}
\newcommand{\OI}{{[O\,{\sc i}]}}
\newcommand{\OIs}{{[O\,{\sc i}]\,}}
\newcommand{\OIl}{{[O\,{\sc i}]\,$\lambda$}}

\newcommand{\CIIs}{{[C\,{\sc ii}]\,}}

\newcommand{\Ha}{H$\alpha$}
\newcommand{\Has}{H$\alpha$\,}
\newcommand{\Hb}{H$\beta$}
\newcommand{\Hbs}{H$\beta$\,}
\newcommand{\zeight}{$z\sim8$\,}
\newcommand{\zsix}{$z\sim6$\,}
\newcommand{\grating}{G395M/F290LP}
\newcommand{\ppxf}{{\sc ppxf}\xspace} 
\begin{document}

   \title{JADES: Probing interstellar medium conditions at $z\sim5.5-9.5$ with ultra-deep \emph{JWST}/NIRSpec spectroscopy}
   \titlerunning{ISM conditions from deep $z\gtrsim6$ spectra}

   \author{
Alex J. Cameron
\inst{1}\thanks{E-mail: alex.cameron@physics.ox.ac.uk}
\and
Aayush Saxena
\inst{1}\fnmsep\inst{2}
\and
Andrew J. Bunker
\inst{1}
\and
Francesco D'Eugenio
\inst{3}\fnmsep\inst{4}
\and
Stefano Carniani
\inst{5}
\and
Roberto Maiolino
\inst{3}\fnmsep\inst{4}\fnmsep\inst{2}
\and
Emma Curtis-Lake
\inst{6}
\and
Pierre Ferruit
\inst{7}
\and
Peter Jakobsen
\inst{8}\fnmsep\inst{9}
\and
Santiago Arribas
\inst{10}
\and
Nina Bonaventura
\inst{8}\fnmsep\inst{9}
\and
Stephane Charlot
\inst{11}
\and
Jacopo Chevallard
\inst{1}
\and
Mirko Curti
\inst{3}\fnmsep\inst{4}\fnmsep\inst{12}
\and
Tobias J. Looser
\inst{3}\fnmsep\inst{4}
\and
Michael V. Maseda
\inst{13}
\and
Tim Rawle
\inst{14}
\and
Bruno Rodríguez Del Pino
\inst{10}
\and
Renske Smit
\inst{15}
\and
Hannah \"Ubler
\inst{3}\fnmsep\inst{4}
\and
Chris Willott
\inst{16}
\and
Joris Witstok
\inst{3}\fnmsep\inst{4}
\and
Eiichi Egami
\inst{17}
\and
Daniel J. Eisenstein
\inst{18}
\and
Benjamin D. Johnson
\inst{18}
\and
Kevin Hainline
\inst{17}
\and
Marcia Rieke
\inst{17}
\and
Brant E. Robertson
\inst{19}
\and
Daniel P. Stark
\inst{17}
\and
Sandro Tacchella
\inst{3}\fnmsep\inst{4}
\and
Christina C. Williams
\inst{20}
\and
Christopher N. A. Willmer
\inst{17}
\and
Rachana Bhatawdekar
\inst{7}\fnmsep\inst{21}
\and
Rebecca Bowler
\inst{22}
\and
Kristan Boyett
\inst{23}\fnmsep\inst{24}
\and
Chiara Circosta
\inst{7}
\and
Jakob M. Helton
\inst{17}
\and
Gareth C. Jones
\inst{1}
\and
Nimisha Kumari
\inst{25}
\and
Zhiyuan Ji
\inst{17}
\and
Erica Nelson
\inst{26}
\and
Eleonora Parlanti
\inst{5}
\and
Lester Sandles
\inst{3}\fnmsep\inst{4}
\and
Jan Scholtz
\inst{3}\fnmsep\inst{4}\fnmsep\inst{2}
\and
Fengwu Sun
\inst{17}
}
\institute{
Department of Physics, University of Oxford, Denys Wilkinson Building, Keble Road, Oxford OX1 3RH, UK\\
\and
Department of Physics and Astronomy, University College London, Gower Street, London WC1E 6BT, UK\\
\and
Kavli Institute for Cosmology, University of Cambridge, Madingley Road, Cambridge CB3 0HA, UK\\
\and
Cavendish Laboratory, University of Cambridge, 19 JJ Thomson Avenue, Cambridge CB3 0HE, UK\\
\and
Scuola Normale Superiore, Piazza dei Cavalieri 7, I-56126 Pisa, Italy\\
\and
Centre for Astrophysics Research, Department of Physics, Astronomy and Mathematics, University of Hertfordshire, Hatfield AL10 9AB, UK\\
\and
European Space Agency (ESA), European Space Astronomy Centre (ESAC), Camino Bajo del Castillo s/n, 28692 Villanueva de la Cañada, Madrid, Spain
\\
\and
Cosmic Dawn Center (DAWN), Copenhagen, Denmark\\
\and
Niels Bohr Institute, University of Copenhagen, Jagtvej 128, DK-2200, Copenhagen, Denmark\\
\and
Centro de Astrobiolog'ia (CAB), CSIC–INTA, Cra. de Ajalvir Km.~4, 28850- Torrej'on de Ardoz, Madrid, Spain\\
\and
Sorbonne Universit\'e, CNRS, UMR 7095, Institut d'Astrophysique de Paris, 98 bis bd Arago, 75014 Paris, France\\
\and
European Southern Observatory, Karl-Schwarzschild-Strasse 2, D-85748 Garching bei Muenchen, Germany\\
\and
Department of Astronomy, University of Wisconsin-Madison, 475 N. Charter St., Madison, WI 53706, USA\\
\and
European Space Agency, Space Telescope Science Institute, Baltimore, Maryland, US\\
\and
Astrophysics Research Institute, Liverpool John Moores University, 146 Brownlow Hill, Liverpool L3 5RF, UK\\
\and
NRC Herzberg, 5071 West Saanich Rd, Victoria, BC V9E 2E7, Canada\\
\and
Steward Observatory, University of Arizona, 933 N. Cherry Ave., Tucson, AZ 85721, USA\\
\and
Center for Astrophysics $|$ Harvard \& Smithsonian, 60 Garden St., Cambridge MA 02138 USA\\
\and
Department of Astronomy and Astrophysics, University of California, Santa Cruz, 1156 High Street, Santa Cruz, CA 95064, USA\\
\and
NSF’s National Optical-Infrared Astronomy Research Laboratory, 950 North Cherry Avenue, Tucson, AZ 85719, USA\\
\and
European Space Agency, ESA/ESTEC, Keplerlaan 1, 2201 AZ Noordwijk, NL\\
\and
Jodrell Bank Centre for Astrophysics, Department of Physics and Astronomy, School of Natural Sciences, The University of Manchester, Manchester, M13 9PL, UK\\
\and
School of Physics, University of Melbourne, Parkville 3010, VIC, Australia\\
\and
ARC Centre of Excellence for All Sky Astrophysics in 3 Dimensions (ASTRO 3D), Australia\\
\and
AURA for European Space Agency, Space Telescope Science Institute, 3700 San Martin Drive. Baltimore, MD, 21210\\
\and
Department for Astrophysical and Planetary Science, University of Colorado, Boulder, CO 80309, USA\\
}

\authorrunning{A. J. Cameron et al. }


 
  \abstract{We present emission line ratios from a sample of 27 Lyman break galaxies from $z\sim5.5-9.5$ with $-17.0 < M_{1500} < -20.4$, measured from ultra-deep \emph{JWST}/NIRSpec MSA spectroscopy from the \textit{JWST Advanced Deep Extragalactic Survey} (JADES).
We use a combination of 28 hour deep PRISM/CLEAR and 7 hour deep G395M/F290LP observations to measure, or place strong constraints on, ratios of widely studied rest-frame optical emission lines including H$\alpha$, H$\beta$, \OIIll 3726, 3729, \NeIIIl 3869, \OIIIl 4959, \OIIIl 5007, \OIl 6300, \NIIl 6583, and \SIIll 6716, 6731 in individual $z>5.5$ spectra.
We find that the emission line ratios exhibited by these $z\sim5.5-9.5$ galaxies occupy clearly distinct regions of line-ratio space compared to typical $z\sim0-3$ galaxies, instead being more consistent with extreme populations of lower-redshift galaxies. This is best illustrated by the \OIII/\OIIs ratio, tracing interstellar medium (ISM) ionisation, in which we observe more than half of our sample to have \OIII/\OIIs $>10$.
Our high signal-to-noise spectra reveal more than an order of magnitude of scatter in line ratios such as \OII/H$\beta$ and \OIII/\OII, indicating significant diversity in the ISM conditions within the sample.
We find no convincing detections of \NIIl 6583 in our sample, either in individual galaxies, or a stack of all \grating\, spectra.
The emission line ratios observed in our sample are generally consistent with galaxies with extremely high ionisation parameters (log $U\sim-1.5$), and a range of metallicities spanning from $\sim0.1\times Z_\odot$ to higher than $\sim0.3\times Z_\odot$, suggesting we are probing low-metallicity systems undergoing periods of rapid star-formation, driving strong radiation fields.
These results highlight the value of deep observations in constraining the properties of individual galaxies, and hence probing diversity within galaxy population.}

   \keywords{galaxies: evolution -- galaxies: high-redshift -- galaxies: ISM}

   \maketitle
%

\section{Introduction} \label{sec:intro}

Emission lines can be some of the most prominent features of galaxy spectra.
Regions of photoionised gas around young stars (H{\sc ii} regions) routinely exhibit bright emission from hydrogen and helium recombination, as well as collisionally excited emission from numerous metal ions, especially O$^{+}$, O$^{2+}$, Ne$^{2+}$, S$^{+}$, N$^{+}$ \citep[e.g.][]{Peimbert2017}.
Ratios of emission line fluxes are sensitive to abundances of metals and their ionisation states, but also the physical conditions of the emitting nebulae, such as temperature and density, which are in turn linked to the nature of the ionising sources powering the emission \citep[e.g.][]{OsterbrockFerland2006}.
Emission line ratios therefore have great diagnostic power for understanding properties of the interstellar medium (ISM) and stellar populations in galaxies (see \citealt{Kewley2019} for a review).

However, observed emission lines in galaxies can arise from other phenomena, including active galacitc nuclei (AGN; \citealt{Kewley2006, Feltre2016}) or shock heated gas \citep{Dopita1996, Allen2008, Sutherland2017}.
Meanwhile, stellar populations other than the young O and B stars typically associated with H{\sc ii} regions, such as high-mass X-ray binaries (HMXBs; \citealt{Senchyna2020}) or post AGB stars \citep{Byler2019}, can power emission in photoionised regions of gas.
Each of these heating sources yields characteristically different emission spectra.
An observed global emission spectrum of a star-forming galaxy will typically be a superposition of emission from many H{\sc ii} regions with different conditions as well as possible contributions from these additional sources, making the interpretation of emission spectra challenging.

Despite this complexity, galaxies have been observed to follow remarkably tight trends in emission line-ratio space.
The so-called BPT and VO87 diagrams, which relate the ratio of \OIIIl 5007/H$\beta$ to ratios of either \NIIl 6583/\Ha, \SIIll 6716,6731/\Ha, or \OIl 6300/\Has \citep{BPT1981, VO1987}, are perhaps the best known examples of such trends\footnote{Diagrams relating \OIII/H$\beta$ to \SII/\Has or \OI/\Has are often referred to as the \SII-BPT and \OI-BPT diagrams. However, they were first introduced by \citet{VO1987}. Throughout this paper we refer to these as VO87 diagrams.}.
With large surveys like SDSS \citep{York2000}, star-forming galaxies at $z\sim0$ have been observed to form a tight sequence in the \NII-BPT diagram, with a general anti-correlation between the two ratios \citep[e.g.][]{Kauffmann2003, Kewley2006}.
This tight correlation is the result of the fact that the two parameters that most strongly dictate the emission spectra of galaxies, metallicity and ionisation parameter\footnote{The dimensional ionisation parameter $q$ is defined as the ratio of ionising flux to hydrogen density (i.e. $q=\Phi/n_H$). Here we refer to the commonly used dimensionless ionisation parameter $U$, which is related to this as $U=q/c$} ($U$), are themselves generally correlated in galaxies \citep{Dopita1986, Carton2017}.
Position within this sequence is thus dominated by these two parameters, although variations in other properties including N/O abundance ratio, electron density and radiation hardness also contribute \citep[e.g.][]{Kewley2013_theory, Bian2016, Curti2022_BPT}.
These correlations are not unique to the BPT and VO87 diagrams, and indeed other combinations of emission lines, such as (\OII+\OIII)/H$\beta$ vs. \OIII/\OIIs (`R23-O32'), have been observed to exhibit similar trends \citep[e.g.][]{VO1987, Kewley2006}.

Existing studies of earlier cosmic epochs show that, compared to the loci inhabited by local galaxies in line-ratio space, $z\sim2$ galaxies are generally offset to higher excitation and higher ionisation \citep[e.g.][]{Strom2017, Runco2021}.
This is perhaps best exhibited by the widely studied offset in the \NII-BPT diagram \citep{Kewley2013_evolution, Steidel2014, Steidel2016, Runco2022}, where $z\sim2$ galaxies are observed with typically higher \NII/\Has at a fixed \OIII/\Hb.
This is often explained as a combination of higher ionisation parameters, higher densities, and harder ionising spectra at high redshift \citep{Brinchmann2008, Kewley2013_theory, Steidel2016, Sanders2016, Hirschmann2017}. Many authors have suggested that super solar $\alpha$/Fe abundance ratios at these earlier times play an important role by driving harder stellar ionisation fields \citep[e.g.][]{Strom2017, Topping2020a, Topping2020b}.
However, the N/O abundance ratio is clearly another important parameter to consider, with some studies suggesting this offset can be explained as the result of higher N/O at fixed \OIII/\Hbs \citep{Masters2016}.

Despite the wealth of rest-frame optical studies highlighting the different ISM conditions in galaxies at $z\sim2$,
emission line studies of galaxies at $z\gtrsim6$ were, until recently, very limited. 
Broad band excesses had already suggested an abundance of galaxies at this epoch with very high equivalent width emission from \OIII +H$\beta$ \citep[e.g.][]{Endsley2021, Endsley2022, Fujimoto2023}. However constraints on their ISM conditions were limited to studies of far-infrared lines, such as \OIIIs88 $\mu$m and \CIIs158 $\mu$m from ALMA, which suggested high ionisation parameters and enhanced $\alpha$/Fe abundance ratios \citep[e.g.][]{Hasimoto2019, Carniani2020, Harikane2020, Fujimoto2022, Witstok2022}. 

The recent commissioning of \emph{JWST} is transforming our ability to probe ISM physics in the early Universe.
Already, emission line measurements from medium-depth observations from the initial months of \emph{JWST} operations have observed $z\gtrsim6$ galaxies to exhibit generally low metallicity, high ionisation, high excitation, and high temperature \citep{Curti2023_ero, Katz2023_ero, FengwuSun2022b, Rhoads2022, Tacchella2022, Trump2022, Langeroodi2022, Williams2022, Heintz2022, Matthee2022_eiger, Fujimoto2023, Mascia2023, Sanders2023, Nakajima2023, Tang2023}.
Furthermore, the observed emission spectra of some of these targets have been found to be consistent with the presence of AGN activity \citep{Brinchmann2022}, while \citet{Katz2023_ero} suggested that the presence of high mass X-ray binaries could help to explain the observed emission line ratios.

Although much insight has already been gained from these early observations, they have largely been based on relatively shallow spectroscopy.
Here we leverage data taken from the `Deep' spectroscopic tier of the \textit{JWST Advanced Deep Extragalactic Survey} (JADES), the deepest spectroscopic observations yet taken with NIRSpec, to provide a more detailed look at the emission line ratios of galaxies at $z\sim5.5-9.5$.
These MSA observations reached exposure times of up to 28 hours in the PRISM/CLEAR ($R\sim100$) and up to 7 hours in each of the 3 medium resolution gratings ($R\sim1000$), providing unprecedented new insights into the ISM properties of galaxies within the first Gyr of the Universe's history.
The structure of this paper is as follows. In Section~\ref{sec:data} we describe the observations, data reduction and measurement of emission line fluxes. In Section~\ref{sec:results} we present our emission line ratios on various diagnostic diagrams and provide comparisons with literature samples.
Section~\ref{sec:discussion} then presents a discussion of the implications of these observations. We summarise in Section~\ref{sec:conclusion}.
Throughout this paper we adopt the \citet{Planck18} cosmology: $\Omega_\Lambda=0.6847$, $\Omega_m=0.3153$, $H_0=67.36$ km s$^{-1}$ Mpc$^{-1}$. All magnitues are quoted in the $AB$ magnitude system \citep{OkeGunn1983}.

\section{Data \& Analysis} \label{sec:data}

\begin{figure*}
    \includegraphics[width=0.98\textwidth]{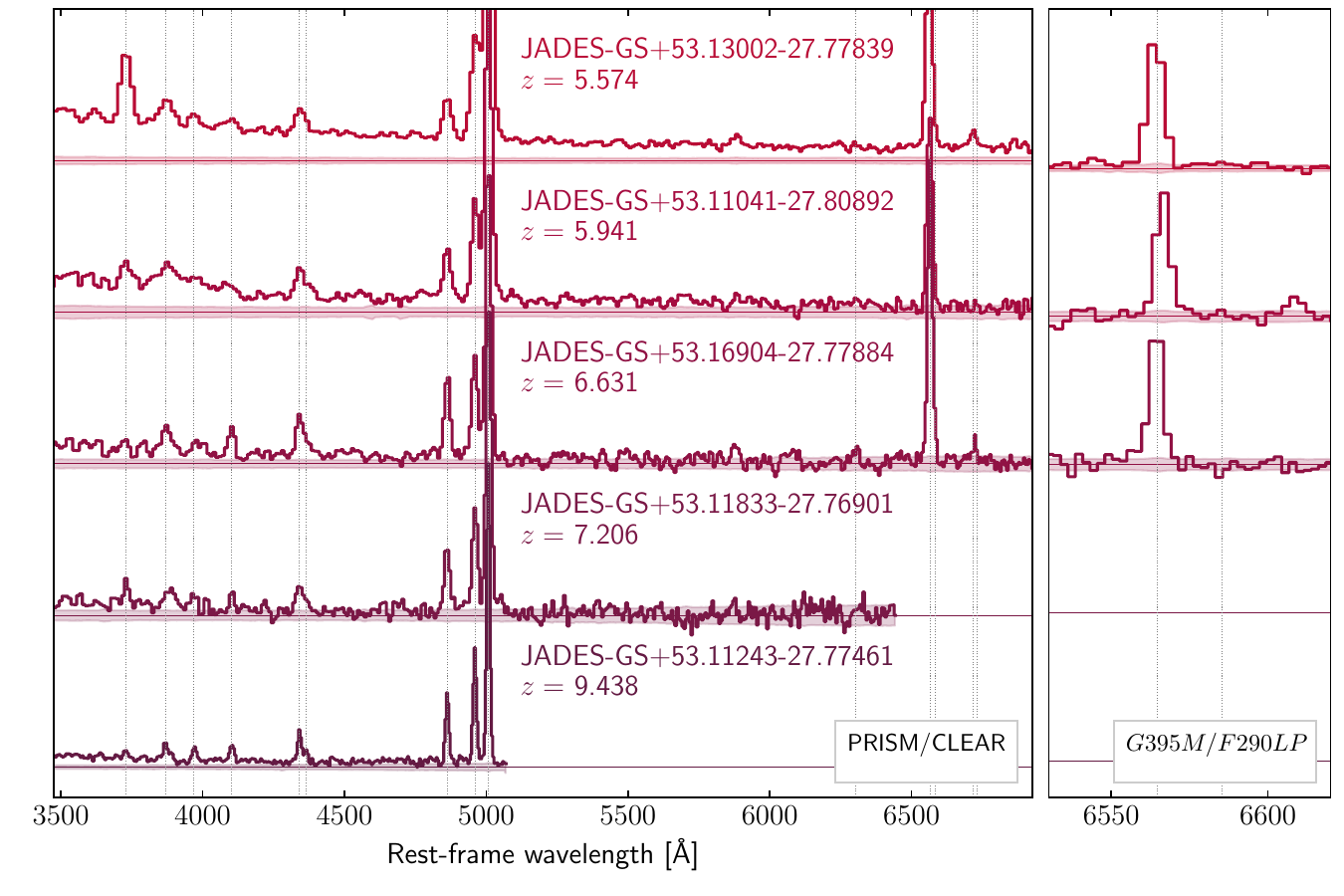}
    \caption{Example spectra from five galaxies included in our sample.
    \textit{Left:} Low-resolution PRISM/CLEAR spectra shifted to the rest-frame according to the observed redshift. Flux is shown in $f_{\lambda}$ but has been renormalised relative to the peak flux from the \OIIIl 5007 line. Solid horizontal lines show the zero flux of each spectrum. The shaded region around the `zero' line indicates the 1-$\sigma$ noise spectrum in the same renormalised units. We show only the subset of spectral coverage that includes the key emission lines used in this study. Vertical dotted lines show the expected centroids of the \OIIl 3727, \NeIIIl 3869, \NeIIIl 3967, H$\delta$, H$\gamma$, \OIIIl 4363, H$\beta$, \OIIIl 4959, \OIIIl 5007, \OIl 6300, H$\alpha$, \NIIl 6583, \SIIl 6716, and \SIIl 6731 lines respectively (moving left to right).
    \textit{Right:} Zoom in on the \Ha+\NIIs complex as observed in the G395M/F290LP grating spectra of these same galaxies. Observed spectra are shown in the same way as for the left panel (except renormalised to H$\alpha$, rather than \OIII). Vertical dotted lines show the expected centroids of H$\alpha$ and \NIIl 6583.
    }
    \label{fig:example_spectra}
\end{figure*}

\subsection{Observations}
\label{sub:observations}

The data presented in this paper were obtained via multi-object spectroscopic observations from \emph{JWST}/NIRSpec using the micro-shutter assembly (MSA; \citealt{Jakobsen2022_NS_Overview, Ferruit2022_NS_MOS}).
Observations were carried out as part of JADES in three visits between 21-25 Oct 2022 (Program ID: 1210; PI: N. Luetzgendorf) in the Great Observatories Origins Deep Survey South (GOODS-S) legacy field in a region overlapping with the Hubble Ultra Deep Field (HUDF; \citealt{Koekemoer2013}).
Each visit consisted of 33,613 s integration in the PRISM/CLEAR low-resolution setting and 8,403 s integration in each of G140M/F070LP, G235M/F170LP, G395M/F290LP, and G395H/F290LP filter/grating settings.
Across three visits, this totals 28 hours of integration in the PRISM, which provides continuous spectral coverage from 0.6 - 5.3 $\mu$m at $R\sim30-300$, and $\sim$7 hours in each of the medium resolution gratings, which combine to provide $R\sim700-1300$ across the full spectral range of NIRSpec, plus 7 hours in the high-resolution grating which provides $R\sim2700$ from $\sim$2.8 - 5.1 $\mu$m.
In this paper we present only measurements from the PRISM/CLEAR and G395M/F290LP observations. Results obtained from other configurations will be presented in forthcoming work.

Observations within each visit were performed as a 3-shutter nod.
The central pointing of each visit was dithered (by $<1$ arcsec) such that common targets were observed in different shutters and different detector real-estate.
Thus, each visit had a unique MSA configuration, although target allocation (performed with the eMPT\footnote{\url{https://github.com/esdc-esac-esa-int/eMPT_v1}}) was optimised for maximising target commonality between all three dither positions.
A total of 253 unique targets were observed in the PRISM configuration with the three dithers featuring 145, 155, and 149 targets respectively. 67 targets were triply exposed in the PRISM and reached full depth, while 62 targets featured in only two dithers and a further 124 received only one-third total depth. The hierarchical allocation of shutters means that the double- and triple-exposed targets are biased toward the highest priority (and hence highest redshift) targets.
Indeed, of the 27 galaxies presented in this work, 18 received triple coverage, 6 received double, and 3 received only single coverage.
Details of the target selection scheme are presented in \citet{Bunker2023_DR}.

The low dispersion associated with the PRISM mode means that all targets are observed with non-overlapping spectra. However, in the higher resolution modes, individual spectra are dispersed over a much larger extent of detector real-estate. To minimise the possibility of contaminating emission, in our grating observations we isolate our highest priority targets by closing the shutters of low-priority targets on the same row (i.e. targets that could cause overlapping spectra).
Thus, for our grating spectra we observe only 198 unique targets (119, 121, and 111 in each dither). However, the high-priority $z\sim5.5-9.5$ galaxies presented in this work are almost unaffected by this measure with only two exceptions. One target (JADES-GS+53.11351-27.77284) that was triply exposed in PRISM was only singly exposed in the gratings, while one target (JADES-GS+53.17582-27.77446) was not observed in the medium resolution setting at all.
The total integration times obtained on each target are summarised in Table~\ref{tab:integration_times}.

\begin{table*}
\centering
\begin{tabular}{cccccccccccccc}
\hline
\hline
ID & NIRSpec\_ID & R.A. & Decl. & $z_{\rm PRISM}$ & $t_{\rm int.}$ (hr) & $t_{\rm int.}$ (hr) & $M_{1500}$ \\
 & & & & & PRISM & $G395M$ &   \\
\hline
\hline
JADES-GS+53.11243-27.77461 & 10058975 & 53.112434 & -27.774609 & 9.438 & 28.0 & 7.0$^a$ & $-20.36\pm0.01$ \\
JADES-GS+53.16446-27.80218 & 8013 & 53.164464 & -27.802180 & 8.479 & 28.0 & 7.0$^a$ & $-17.55\pm0.84$ \\
JADES-GS+53.15682-27.76716 & 21842 & 53.156825 & -27.767159 & 7.981 & 28.0 & 7.0$^a$ & $-18.43\pm0.03$ \\
JADES-GS+53.16746-27.77201 & 10013682 & 53.167464 & -27.772006 & 7.277 & 18.7 & 4.7$^a$ & $-17.01\pm0.22$ \\
JADES-GS+53.11833-27.76901 & 10013905 & 53.118327 & -27.769010 & 7.206 & 28.0 & 7.0$^a$ & $-18.88\pm0.05$ \\
JADES-GS+53.13423-27.76891 & 20961 & 53.134229 & -27.768913 & 7.051 & 18.7 & 4.7$^a$ & $-18.82\pm0.06$ \\
\hline
JADES-GS+53.11730-27.76408 & 10013609 & 53.117303 & -27.764082 & 6.931 & 9.3 & 2.3 & $-18.71\pm0.06$ \\
JADES-GS+53.15579-27.81520 & 4297 & 53.155788 & -27.815202 & 6.718 & 9.3 & 2.3 & $-18.58\pm0.07$ \\
JADES-GS+53.15138-27.81917 & 3334 & 53.151381 & -27.819165 & 6.709 & 28.0 & 7.0 & $-17.80\pm0.07$ \\
JADES-GS+53.16904-27.77884 & 16625 & 53.169042 & -27.778838 & 6.631 & 28.0 & 7.0 & $-18.75\pm0.03$ \\
JADES-GS+53.13492-27.77271 & 18846 & 53.134917 & -27.772709 & 6.342 & 28.0 & 7.0 & $-20.14\pm0.01$ \\
JADES-GS+53.17582-27.77446 & 18179 & 53.175816 & -27.774464 & 6.335 & 18.7 & 0.0 & $-18.71\pm0.06$ \\
JADES-GS+53.16660-27.77240 & 18976 & 53.166602 & -27.772403 & 6.329 & 28.0 & 7.0 & $-18.61\pm0.04$ \\
JADES-GS+53.15613-27.77584 & 17566 & 53.156128 & -27.775841 & 6.105 & 18.7 & 4.7 & $-19.05\pm0.04$ \\
JADES-GS+53.16062-27.77161 & 19342 & 53.160620 & -27.771613 & 5.981 & 28.0 & 7.0 & $-18.67\pm0.04$ \\
JADES-GS+53.11911-27.76080 & 10013618 & 53.119112 & -27.760802 & 5.948 & 28.0 & 7.0 & $-19.42\pm0.02$ \\
JADES-GS+53.12176-27.79763 & 9422 & 53.121755 & -27.797634 & 5.943 & 28.0 & 7.0$^a$ & $-19.81\pm0.01$ \\
JADES-GS+53.11041-27.80892 & 6002 & 53.110411 & -27.808923 & 5.941 & 28.0 & 7.0 & $-18.60\pm0.03$ \\
JADES-GS+53.12259-27.76057 & 10013620 & 53.122590 & -27.760569 & 5.920 & 28.0 & 7.0 & $-19.64\pm0.02$ \\
JADES-GS+53.17655-27.77111 & 19606 & 53.176550 & -27.771108 & 5.891 & 9.3 & 2.3 & $-18.75\pm0.06$ \\
JADES-GS+53.11351-27.77284 & 10056849 & 53.113511 & -27.772836 & 5.822 & 28.0 & 2.3 & $-18.16\pm0.05$ \\
JADES-GS+53.16730-27.80287 & 10005113 & 53.167303 & -27.802870 & 5.820 & 28.0 & 7.0$^a$ & $-17.85\pm0.08$ \\
JADES-GS+53.15407-27.76607 & 22251 & 53.154070 & -27.766072 & 5.804 & 18.7 & 4.7 & $-18.93\pm0.04$ \\
JADES-GS+53.11537-27.81477 & 4404 & 53.115372 & -27.814771 & 5.775 & 28.0 & 7.0 & $-19.31\pm0.03$ \\
JADES-GS+53.13002-27.77839 & 16745 & 53.130023 & -27.778393 & 5.574 & 28.0 & 7.0 & $-19.60\pm0.01$ \\
JADES-GS+53.12972-27.80818 & 6246 & 53.129722 & -27.808177 & 5.570 & 28.0 & 7.0 & $-18.01\pm0.05$ \\
JADES-GS+53.11572-27.77496 & 10016374 & 53.115717 & -27.774955 & 5.507 & 18.7 & 4.7 & $-18.66\pm0.02$ \\
\hline
\hline
\end{tabular}
\caption{
Summary of total integration times on each of the targets presented in this sample. The six targets above the horizontal rule comprise the JADES `$z\sim8$' sample and do not have coverage of any of \Ha, \NIIs or \SIIs in either PRISM or \grating\, data. The remainder are referred to as the `$z\sim6$' sample. `NIRSpec\_ID' refers to the integer ID used in the JADES data release \citep{Bunker2023_DR}. \\
$^{a}$ Target does not have coverage of the \Ha+\NIIs complex in \grating\, observation.  \\
}
\label{tab:integration_times}
\end{table*}

\subsection{Data reduction}
\label{sub:data_reduction}

These observations were processed by adopting algorithms developed by the ESA NIRSpec Science Operations Team (SOT) and the NIRSpec GTO Team. Details of the data-processing workflow will be presented in a forthcoming NIRSpec GTO collaboration paper. As this work uses the observations carried out with only two spectral configurations of the program, we describe the main steps for the PRISM/CLEAR and G395M/F290LP filter/grating settings. 
Upon retrieving the level-1a data from the MAST archive, we estimated the count rate per pixel by using the unsaturated groups in the ramp and removing jumps due to cosmic rays identified by estimating the slope of the individual ramps. During this first stage, we also performed the master bias and dark subtraction, corrected snowball artifacts, and flagged saturated pixels.
We then performed the pixel-by-pixel background subtraction by combining the three nod exposures of each pointing. We note that for some targets we exclude one of the 3-shutter nods in the background subtraction stage due to the serendipitous presence of a source contaminating the open background shutter. We then created two-dimensional (2D) cutouts of each 3-shutter slit and performed the flat-field, spectrograph optics, and disperser corrections.
We then ran the absolute calibration and corrected the 2D spectra for path-losses.
Path-loss corrections are dependent on the morphology of the target and its relative position within the shutter.
In this paper, we apply the path-loss corrections assuming a point-like source at the location of the target centroid.
The high-redshift targets presented in this paper are generally smaller, or comparable in size, to the angular resolution of the telescope at the observed wavelength of emission lines used in this study ($\lambda\gtrsim3$ $\mu$m).
We estimate that for the sample presented here, even the ratio with the longest wavelength baseline (\OIIIl 5007 / \OIIll 3726, 3729) would be affected only at the 1 \% (0.01 dex) level by adopting an extended morphology based on fitting to broad-band imaging.
Since it is not \textit{a priori} known how closely the spatial distribution of emission lines trace the starlight observed in broad-band imaging, we decide to adopt the point-like assumption, but acknowledge that this effect introduces some systematic uncertainty to our measurements.

We rectified and interpolated the 2D continuum map onto a regular grid for the G395M/F290LP observation, and an irregular grid for the PRISM/CLEAR to avoid an oversampling of the line spread function at short wavelengths. Finally, the 1D spectra were extracted from the 2D map adopting a box-car aperture as large as the shutter size, centred on the relative position of the target in the shutter. For each target, we combined all 1D spectra and removed any bad pixels with sigma-clipping. 
Cutouts of five example representative 1D spectra are shown in Figure~\ref{fig:example_spectra}.

As a verification of our flux calibration, we compared the NIRSpec spectrophotometry to broadband photometry from NIRCam and found the differences to be on average less than 10 \% even before considering the effects of slit-losses due to the MSA. Details of this comparison are presented in Appendix~\ref{sec:flux_cal_verify}.

\subsection{Emission line ratio sample}
\label{sub:sample}

\begin{figure}
    \centering
    \includegraphics[width=0.49\textwidth]{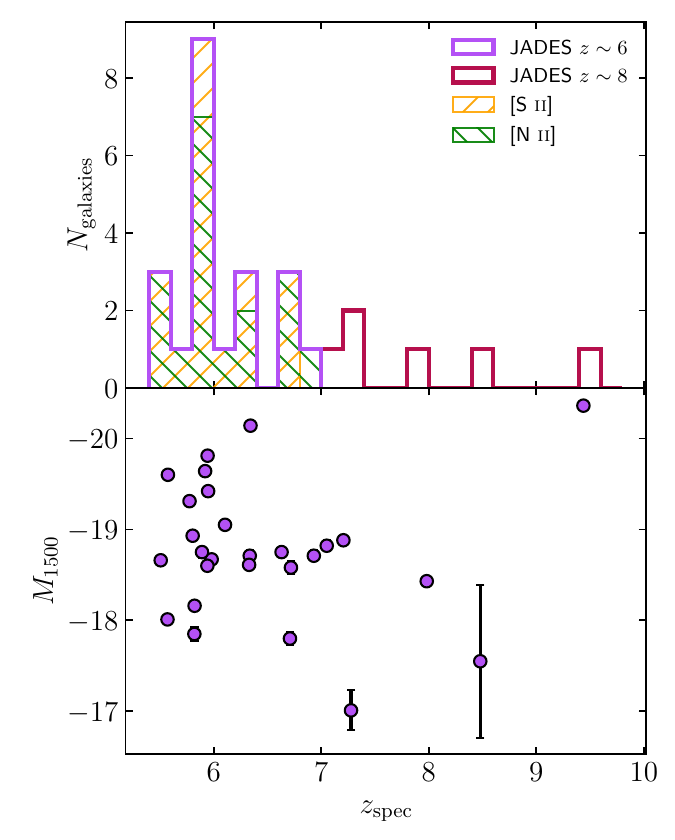}
    \caption{\textit{Top:} Redshift histogram of galaxies observed in this study. We divide the sample into two sub-samples `$z\sim$6' and `$z\sim8$', based simply on cutting at $z=7$ since this is the redshift at which H$\alpha$ is redshifted beyond the spectral coverage of NIRSpec. Thus, for JADES `$z\sim8$' we only have coverage of lines from \OIIll 3726, 3729 to \OIIIl 5007, while for JADES `$z\sim6$' we can in principle observe all the rest-frame optical lines out to \SIIll 6716, 6731.
    The orange hatched histogram indicates galaxies for which we can constrain the \SIIll 6716, 6731 flux. Green hatching is the same but for \NIIl 6583.
    \textit{Bottom:} Redshift vs. $M_{1500}$ for the full $z\sim5.5-9.5$ sample. We derived $M_{1500}$ directly from the PRISM spectra as described in Section~\ref{sub:muv_calc}.
    }
    \label{fig:sample_hist}
\end{figure}

Although 253 unique targets were observed in these observations, this study focuses only on a rest-UV selected sample of confirmed $z>5.5$ galaxies with robustly detected emission lines.
The criteria used to select targets for this program are described in detail in \citet{Bunker2023_DR}. We describe here only the aspects most relevant to this work.

The highest priorities in target assignment were reserved for galaxies at the highest redshifts, with a preference for the UV-bright galaxies. 
Candidate $z>5.5$ galaxies were primarily drawn from a compilation of existing literature catalogs based on Hubble Space Telescope (\emph{HST}) broad-band imaging \citep[e.g][]{Bunker2004, Bouwens2015, Bouwens2021, Finkelstein2015, Harikane2016}, with a number of the highest redshift galaxies presented here having been identified with \emph{HST} \citep[e.g.][]{Bouwens2011, Bouwens2016, Ellis2013, McLure2013, Oesch2013}. In fact, only five galaxies presented in this work were newly identified from \emph{JWST}/NIRCam imaging (see \citealt{Bunker2023_DR} for details).
In total, 47 galaxies were assigned MSA shutters as $z>5.7$ candidates out of 299 possible such candidates within the MSA footprint. Of these, 28 passed redshift-dependent rest-UV cuts to place them in our highest priority classes $PC=1$ and $PC=4$ ($PC=2$ and 3 contained only 5 targets between them, none of which featured here; see \citealt{Bunker2023_DR} for class descriptions).
For this analysis, we consider only galaxies with robust spectroscopically confirmed redshifts in which H$\beta$ is detected with $S/N>5$. Of the 28 rest-UV-selected targets in $PC=1$ and $PC=4$, 18 met these criteria. 
We also include 6 (1) galaxies selected as $z>5.7$ candidates that had failed the rest-UV selection cut, but were included in $PC=6.1$ ($PC=6.2$) as a result of meeting the magnitude criterion of $m_{\rm F160W}<29$ ($m_{\rm F160W}>29$) and turned out to match our H$\beta$ $S/N$ criterion for this paper.
Finally, three galaxies selected in $PC=7.5$ ($4.5<z<5.7$, $m_{\rm F160W}<29$) had $z_{\rm spec}>5.5$ and $S/N_{H\beta}>5$, and we included these in our sample as well.
One galaxy exhibited a broad-component under \Has in $R\sim1000$ spectroscopy, suggestive of nuclear activity, and was discarded so as to avoid contamination from emission from AGN.
Thus, our final sample included 27 star-forming galaxies, drawn from a largely rest-UV-selected sample.

This work focuses predominantly on the low-resolution PRISM/CLEAR spectra (`PRISM' hereafter), since these are significantly deeper than our medium-resolution grating spectra.
The spectral resolution of the PRISM varies significantly with wavelength from $R\sim30$ around 1.2 $\mu$m up to $R\sim300$ at the long wavelength cutoff around 5 $\mu$m (e.g. Figure 5 in \citealt{Jakobsen2022_NS_Overview}).
By limiting ourselves to rest-frame optical emission lines at $z\gtrsim5.5$, the measurements presented in this paper are made at observed wavelengths of $\lambda_{\rm obs} \gtrsim 2.4$ $\mu$m where the resolution is $R\gtrsim100$, and improving steeply as a function of wavelength.
Thus, in all cases from the PRISM, the \OIIIs $\lambda\lambda$4959, 5007 doublet is completely resolved from \Hb. The doublet itself is always at least partially resolved, and is completely resolved in our highest redshift targets (Figure~\ref{fig:example_spectra}).
\OIIll 3726, 3729 and \SIIll 6716, 6731 are never resolved into two components, but the latter can easily be fully resolved from the \Ha+\NIIs complex over this redshift range.
However, \Has and \NIIl 6583 require resolutions of $R\sim500$ to be fully resolved and thus are significantly blended across the full redshift range.
Constraints on this ratio are instead derived from our higher resolution (but shallower) \grating\, grating spectra.

Figure~\ref{fig:sample_hist} shows the redshift histogram of galaxies in this sample.
We divide galaxies into two subsamples, based on the observability of \Ha.
The purple `$z\sim6$' sample spans over $5.507\leq z \leq 6.931$ ($z_{\rm median} = 5.943$) and includes 21 galaxies for which \Has is still within the wavelength coverage of NIRSpec.
There are an additional six galaxies in our `$z\sim8$' sample, spanning over $7.05\leq z \leq 9.45$ ($z_{\rm median} = 7.629$), for which we study only emission lines from \OIIll 3726, 3729 to \OIIIl 5007.
Since the \grating\, spectra are dispersed across a larger area of detector real estate, they do not provide continuous spectral coverage over the nominal spectral range. For 2/21 galaxies in the \zsix sample, we do not have \NII/\Has constraints because these lines fall into the detector gap of NIRSpec. We additionally have one galaxy (JADES-GS+53.17582-27.77446) which was only observed in the PRISM mode (see Section~\ref{sub:observations}).
For one galaxy in the \zsix sample, the \SIIs doublet falls beyond the observed spectral range.

\subsection{Spectral analysis and line flux measurements}
\label{sub:spectral_fitting}

\subsubsection{Calculation of $M_{1500}$}
\label{sub:muv_calc}

The continuum is generally well detected in the rest-frame ultraviolet (UV) across our sample. In all except our faintest target, the signal-to-noise on the continuum reaches $S/N>3$ for individual spectral pixels long-ward of the Lyman break, and in six spectra we reach in excess of $S/N\gtrsim10$ per pixel in the rest-frame UV.
For our faintest target, JADES-GS+53.16746-27.77201, we do observe faint continuum in the rest-UV, but the signal-to-noise per pixel is only $S/N\approx1.8$.
We derive rest-frame UV absolute magnitudes ($M_{1500}$) for our sample directly from the spectra. We shift the observed spectra to the rest-frame using the spectroscopically confirmed redshift and integrate across a synthetic 200~\AA{} wide top-hat filter centred on 1500~\AA{}. This integrated flux is then converted into an absolute magnitude.
Derived $M_{1500}$ values are reported in Table~\ref{tab:integration_times}.
Aside from our faintest target (JADES-GS+53.16746-27.77201 with $M_{1500}=-17.01\pm0.22$), all other targets have magnitudes in the range $-17.5 \leq M_{1500} \leq -20.4$.

\subsubsection{PRISM/CLEAR emission line measurements}

Emission line measurements and continuum modelling are performed simultaneously using the penalised pixel fitting algorithm, \ppxf
\citep{Cappellari2017, Cappellari2022}. \ppxf models the continuum as a linear superposition of simple stellar-population (SSP) spectra, using
non-negative weights and matching the spectral resolution of the observed spectrum. As input, we used the high-resolution (R=10,000) SSP library
combining MIST isochrones \citep{Choi+2016} and the C3K theoretical atmospheres \citep{Conroy+2019}. The flux blue-ward of the Lyman break was manually set to 0.
These templates are complemented by a 5\textsuperscript{th}-degree multiplicative Legendre polynomial, to take into account systematic differences
between the SSPs and the data (e.g., dust, mismatch between the SSP models and high-redshift stellar populations, and residual flux calibration problems). 
The emission lines are modelled as pixel-integrated Gaussians, again matching the observed spectral resolution. To reduce the number of degrees
of freedom, we divide the emission lines into four kinematic groups, constrained to have the same redshift and \emph{intrinsic} broadening. These groups are: UV lines (blueward of 3000~\AA\ in the rest-frame), the Balmer series of Hydrogen, non-Hydrogen optical lines (blueward of 9000~\AA\ in the rest-frame), and near-infrared (NIR) lines. The stellar continuum has
the same kinematics as the Balmer lines.
Furthermore, we tie together doublets that have fixed ratios, and constrain variable-ratio doublets to their physical ranges. 
In particular, this study focuses on the following lines of interest: \OIIll 3726, 3729, \NeIIIl 3869, 
H$\gamma$, \Hb, \OIIIl 4959, \OIIIl 5007, \OIl 6300, \Ha, \NIIl 6583, \SIIll 6716, 6731.
We note that, at the resolution of the PRISM/CLEAR mode, the two components of the \OIIll 3726, 3729 doublet are significantly blended and so are fit as a single Gaussian component.
We visually inspect these fits to the 1D spectra to ensure they are reliable. For the faintest lines, we also individually inspect the 2D spectrum for each target and confirm that the emission line can be visually identified along the same trace to confirm the measured flux is not affected by uncorrected artifacts.
Where a line is not detected, we derive an upper limit directly from the noise spectrum.
The 1-$\sigma$ upper limit is obtained by integrating the variance spectrum over an interval spanning 1.4$\times\sigma_{\rm line}$ either side of the expected centroid of the line (where $\sigma_{\rm line}$ is the Gaussian line width), and multiplying this by the spectral extent of a pixel.

We note that, at the spectral resolution of our PRISM observations, \NeIIIl 3869 is partially blended with the He {\sc i} $\lambda$3889 line.
From grating spectra, we expect the \NeIIIs line to be brighter than He {\sc i}, typically by at least a factor of two, suggesting that performing a two-component fit can recover the \NeIIIs flux (e.g. Appendix C3 in \citealt{Cameron2021}).
The fitting procedure outlined above simultaneously fits for both lines in this complex, and we confirm from visual inspection that these fits model the complex well in cases with a high signal-to-noise ratio.
In this paper, the only ratio for which we consider the \NeIIIs line is the \NeIIIl 3869 / \OIIll 3726, 3729 ratio. This ratio has a very short wavelength baseline and hence the uncertainties introduced by partial blending are somewhat offset by having almost no dependence on wavelength dependent corrections or calibrations.
Nonetheless, we note that that the \NeIII-based measurements presented in this work should be treated with caution, and that any inference requiring high precision \NeIII-based measurements should be performed with medium- to high-resolution spectra.

\subsubsection{Constraints on \NII/\Has from G395M/F290LP}
\label{sub:grating_fitting}

As discussed in Section~\ref{sub:sample}, the only line ratio for which we consider the data from our higher spectral resolution, but much shallower, \grating\, spectra is \NIIl 6583 / \Ha. These lines are sufficiently close in wavelength that they remain at least partially blended in PRISM/CLEAR spectra across the entire redshift range considered here.
However, in the $R\sim1000$ \grating\, spectra, this complex is always completely resolved.

We performed a visual inspection simultaneously for the PRISM and \grating\, 1D and 2D spectra of each target to identify the possible presence of \NII. In fact, we find no convincing evidence for the detection of \NIIl 6583 in any of our 18 galaxies, either from the resolved $R\sim1000$ spectra, or by way of the appearance of a `red wing' in the \Has profile in the PRISM spectra.
We derive upper-limits on the \NII/\Has ratio in the same way as described above, fitting the \Has line with pPXF, and integrating the variance spectrum across the expected spectral extent of \NIIl 6583, based on the redshift and line-width measured for \Ha.

\subsubsection{Emission line ratio measurements}
\label{sub:line_ratios_measure}

In this paper we focus on the following key diagnostic line ratios:

\begin{multline}
\label{eqn:ratios}
    N2 = \text{log\, \big{(} [N {\sc ii}]} \lambda 6583 / \text{H} \alpha \, \big{)},\\
    S2 = \text{log\, \big{(} [S {\sc ii}]} \lambda\lambda 6716, 6731 / \text{H} \alpha \, \big{)},\\
    O1 = \text{log\, \big{(} [O {\sc i}]} \lambda 6300 / \text{H} \alpha \, \big{)},\\
    R2 = \text{log\, \big{(} [O {\sc ii}]} \lambda\lambda 3726, 3729 / \text{H} \beta \, \big{)},\\
    R3 = \text{log\, \big{(} [O {\sc iii}]} \lambda 5007 / \text{H} \beta \, \big{)},\\
    R23 = \text{log\,} \big{(} \left( \text{[O {\sc iii}]} \lambda\lambda 4959, 5007 + \text{[O {\sc ii}]} \lambda\lambda 3726, 3729 \right) / \text{H} \beta \, \big{)},\\
    O32 = \text{log\, \big{(} [O {\sc iii}]} \lambda 5007 / \text{[O {\sc ii}]} \lambda\lambda 3726, 3729 \, \big{)},\\
    Ne3O2 = \text{log\, \big{(} [Ne {\sc iii}]} \lambda 3869 / \text{[O {\sc ii}]} \lambda\lambda 3726, 3729 \, \big{)}.\\
\end{multline}

Several of these line ratios, such as $N2$, $S2$ and $R3$, are calculated across short wavelength baselines and are thus largely insensitive to dust attenuation or wavelength-dependent calibration issues.
However, to accommodate the analysis of the longer baseline ratios, we correct our emission line fluxes and upper limits for dust extinction according to the SMC dust law from \citet{Gordon2003}. We calculate the $E(B-V)$ from the decrements of Balmer lines. For the \zsix sample where \Has is available, we use the \Ha/\Hbs decrement, assuming an intrinsic ratio of H$\alpha$/H$\beta=2.86$ \citep{OsterbrockFerland2006}. 
For the \zeight sample, where possible, we use H$\gamma$/\Hbs assuming an intrinsic ratio of H$\gamma$/H$\beta=0.468$. Above $z>7$, H$\gamma$ is sufficiently deblended from \OIIIl 4363 so as to avoid a biased measurement. In two cases, the S/N on H$\gamma$ is too low. For these galaxies, we adopt the attenuation obtained from SED fitting with {\sc BEAGLE} (Chevallard et al. in prep.), but find that only very small corrections ($E(B-V)<0.02$) are required.
Overall, the mean $E(B-V)$ from this sample is 0.1.
Our final dust-corrected emission line ratios are presented in Table~\ref{tab:line_ratios}.

\subsection{Stacked spectra}
\label{sub:stacking}

Although this paper focuses on presenting line ratio measurements made for individual objects, we also perform a stack of all the spectra in each of our two redshift bins to aid in assessing median quantities of the sample, and push the limits of emission line non-detections.
In our $z\sim6$ bin, we perform this stacking for both the medium-resolution G395M/F290LP spectra (for the $N2$ ratio) and the low-resolution PRISM spectra (for all other line ratios). In the $z\sim8$ bin we stack only the PRISM spectra.

Each spectrum is de-redshifted to the rest-frame and then re-sampled onto a uniform wavelength grid with spectral pixels 3~\AA{} wide in the rest-frame.
For the $z\sim6$ stack, we renormalise each spectrum by the H$\alpha$ flux, while for the $z\sim8$ stack we renormalise by H$\beta$.
We then take the median across all spectra in each spectral bin to form our final composite spectrum.

In order to extract emission line fluxes, we first subtract off the continuum by masking the regions of bright emission lines and fitting a spline. We then fit each emission line in the continuum-subtracted spectra with single-component Gaussian profiles.
We correct the measured emission line ratios for the effects of dust-reddening in the same way as for the individual measurements, described in Section~\ref{sub:line_ratios_measure}, using the measured \Ha/\Hbs ratio to derive $E(B-V)$ for the $z\sim6$ sample and the H$\gamma$/H$\beta$ ratio for the $z\sim8$ sample.

We repeat the stacking process and line measurements for each sample where each spectrum is instead normalised by the UV continuum flux (taken as the median flux between 1400 and 1600 \AA{} in the rest-frame spectrum). Finally, we repeat these measurements removing the target with the highest average noise level. Although the ratios measurement from these alternative stacks do not vary so much as to change any of the key findings in this paper, the change is larger than the statistical uncertainties on the measurements. 
For the final dust-corrected emission line ratios reported in Table~\ref{tab:stack_values}, we adopt the values measured from the Balmer-emission-normalised stacks. However, rather than adopting the statistical uncertainty, we report the systematic uncertainty, which we take as the maximum difference between the values derived from any two of the stacking regimes described above.

{\renewcommand{\arraystretch}{1.4}%
\begin{landscape}
\begin{table}
\centering
\begin{tabular}{cccccccccccccc}
\hline
\hline
ID & $z_{\rm PRISM}$ & $N2$ & $S2$ & $O1$ & $R3$ & $R2$ & $R23$ & $O32$ & $Ne3O2$ \\
\hline
\hline
JADES-GS+53.11243-27.77461 & 9.438 & ... & ... & ... & $0.62\pm0.01$ & $-0.93\pm0.07$ & $0.75\pm0.01$ & $1.55\pm0.07$ & $0.32\pm0.08$ \\
JADES-GS+53.16446-27.80218 & 8.479 & ... & ... & ... & $0.71\pm0.06$ & $<-0.49$ & $0.84\pm0.06$ & $>1.20$ & ...$^{c}$ \\
JADES-GS+53.15682-27.76716 & 7.981 & ... & ... & ... & $0.77\pm0.02$ & $-0.37\pm0.06$ & $0.92\pm0.02$ & $1.14\pm0.05$ & $0.07\pm0.07$ \\
JADES-GS+53.16746-27.77201 & 7.277 & ... & ... & ... & $0.78\pm0.07$ & $-0.19\pm0.17$ & $0.93\pm0.07$ & $0.97\pm0.16$ & $<-0.04$ \\
JADES-GS+53.11833-27.76901 & 7.206 & ... & ... & ... & $0.64\pm0.06$ & $-0.23\pm0.25$ & $0.81\pm0.07$ & $0.86\pm0.25$ & $-0.15\pm0.39$ \\
JADES-GS+53.13423-27.76891 & 7.051 & ... & ... & $<-0.65$ & $0.53\pm0.04$ & $-0.44\pm0.14$ & $0.69\pm0.04$ & $0.97\pm0.13$ & $<-0.07$ \\
\hline
JADES-GS+53.11730-27.76408 & 6.931 & $<-0.81$ & ... & $<-1.11$ & $0.78\pm0.03$ & $-0.23\pm0.08$ & $0.94\pm0.03$ & $1.01\pm0.07$ & $0.06\pm0.10$ \\
JADES-GS+53.15579-27.81520 & 6.718 & $<-0.75$ & $<-1.01$ & $<-1.12$ & $0.79\pm0.03$ & $-0.78\pm0.23$ & $0.92\pm0.03$ & $1.57\pm0.23$ & $0.48\pm0.24$ \\
JADES-GS+53.15138-27.81917 & 6.709 & $<-0.39$ & $<-0.59$ & $<-0.62$ & $0.94\pm0.13$ & $-0.04\pm0.42$ & $1.10\pm0.14$ & $0.98\pm0.40$ & $<0.36$ \\
JADES-GS+53.16904-27.77884 & 6.631 & $<-1.06$ & $<-1.36$ & $<-1.43$ & $0.54\pm0.01$ & $<-0.87$ & $0.67\pm0.01$ & $>1.41$ & $>0.38$ \\
JADES-GS+53.13492-27.77271 & 6.342 & $<-1.29$ & $<-1.55$ & $<-1.64$ & $0.64\pm0.01$ & $-0.76\pm0.06$ & $0.78\pm0.01$ & $1.40\pm0.05$ & $0.24\pm0.06$ \\
JADES-GS+53.17582-27.77446 & 6.335 & ...$^{b}$ & $<-0.91$ & $<-0.97$ & $0.75\pm0.04$ & $0.18\pm0.06$ & $0.96\pm0.04$ & $0.57\pm0.05$ & $<-0.53$ \\
JADES-GS+53.16660-27.77240 & 6.329 & $<-0.75$ & $<-1.03$ & $<-1.12$ & $0.63\pm0.03$ & $-0.53\pm0.13$ & $0.78\pm0.03$ & $1.15\pm0.13$ & $0.09\pm0.16$ \\
JADES-GS+53.15613-27.77584 & 6.105 & $<-0.80$ & $<-1.02$ & $-1.59\pm0.50$ & $0.73\pm0.04$ & $0.35\pm0.06$ & $0.97\pm0.04$ & $0.37\pm0.04$ & $<-0.68$ \\
JADES-GS+53.16062-27.77161 & 5.981 & $<-0.95$ & $<-1.18$ & $<-1.24$ & $0.65\pm0.02$ & $-0.93\pm0.27$ & $0.78\pm0.02$ & $1.58\pm0.27$ & $<0.22$ \\
JADES-GS+53.11911-27.76080 & 5.948 & $<-0.85$ & $<-1.08$ & $<-1.15$ & $0.68\pm0.03$ & $0.33\pm0.04$ & $0.93\pm0.03$ & $0.35\pm0.03$ & $<-0.71$ \\
JADES-GS+53.12176-27.79763 & 5.943 & ...$^{a}$ & $<-1.87$ & $<-2.02$ & $0.75\pm0.01$ & $-0.82\pm0.04$ & $0.88\pm0.01$ & $1.56\pm0.04$ & $0.36\pm0.04$ \\
JADES-GS+53.11041-27.80892 & 5.941 & $<-1.01$ & $<-1.22$ & $<-1.28$ & $0.74\pm0.02$ & $-0.22\pm0.07$ & $0.90\pm0.02$ & $0.96\pm0.06$ & $-0.01\pm0.09$ \\
JADES-GS+53.12259-27.76057 & 5.920 & $<-0.83$ & $-0.90\pm0.09$ & $<-1.17$ & $0.71\pm0.02$ & $0.04\pm0.04$ & $0.90\pm0.02$ & $0.67\pm0.03$ & $-0.45\pm0.08$ \\
JADES-GS+53.17655-27.77111 & 5.891 & $<-0.96$ & $<-1.05$ & $<-1.08$ & $0.75\pm0.04$ & $-0.23\pm0.12$ & $0.91\pm0.04$ & $0.98\pm0.11$ & $-0.04\pm0.15$ \\
JADES-GS+53.11351-27.77284 & 5.822 & $<-0.80$ & $<-1.29$ & $<-1.35$ & $0.52\pm0.02$ & $-0.61\pm0.10$ & $0.67\pm0.02$ & $1.13\pm0.10$ & $<-0.21$ \\
JADES-GS+53.16730-27.80287 & 5.820 & ...$^{a}$ & $<-1.09$ & $<-1.15$ & $0.53\pm0.03$ & $<-0.55$ & $0.65\pm0.03$ & $>1.08$ & ...$^{c}$ \\
JADES-GS+53.15407-27.76607 & 5.804 & $<-1.34$ & $<-1.53$ & $<-1.54$ & $0.83\pm0.02$ & $-0.14\pm0.05$ & $0.99\pm0.02$ & $0.96\pm0.05$ & $-0.05\pm0.07$ \\
JADES-GS+53.11537-27.81477 & 5.775 & $<-1.13$ & $<-1.38$ & $<-1.45$ & $0.65\pm0.01$ & $-0.40\pm0.05$ & $0.80\pm0.02$ & $1.05\pm0.05$ & $-0.07\pm0.08$ \\
JADES-GS+53.13002-27.77839 & 5.574 & $<-1.15$ & $-1.08\pm0.08$ & $<-1.40$ & $0.82\pm0.03$ & $0.36\pm0.04$ & $1.05\pm0.03$ & $0.46\pm0.03$ & $-0.57\pm0.10$ \\
JADES-GS+53.12972-27.80818 & 5.570 & $<-0.89$ & $-1.17\pm0.13$ & $<-1.23$ & $0.61\pm0.03$ & $-0.01\pm0.06$ & $0.80\pm0.03$ & $0.61\pm0.05$ & $-0.28\pm0.10$ \\
JADES-GS+53.11572-27.77496 & 5.507 & $<-1.17$ & $<-1.42$ & $<-1.46$ & $0.82\pm0.02$ & $-0.11\pm0.04$ & $0.99\pm0.02$ & $0.93\pm0.03$ & $-0.10\pm0.05$ \\
\hline
\hline
\end{tabular}
\caption{Table of dust-corrected emission line ratios measured for each of the 27 galaxies in our sample. Line ratio names are defined in Equation~\ref{eqn:ratios}. Non-detections are quoted as 3-$\sigma$ upper limits. \\
$^{a}$ Target does not have coverage of the \Ha+\NIIs complex in \grating\, observation. \\
$^{b}$ Target was not observed in our \grating\, MSA configuration. \\
$^{c}$ $Ne3O2$ ratio is unconstrained as neither \OIIs nor \NeIIIs were robustly detected.
}
\label{tab:line_ratios}
\end{table}
\end{landscape}
}

\section{Diagnostic diagrams} 
\label{sec:results}

\begin{figure*}
    \centering
    \includegraphics[width=\textwidth]{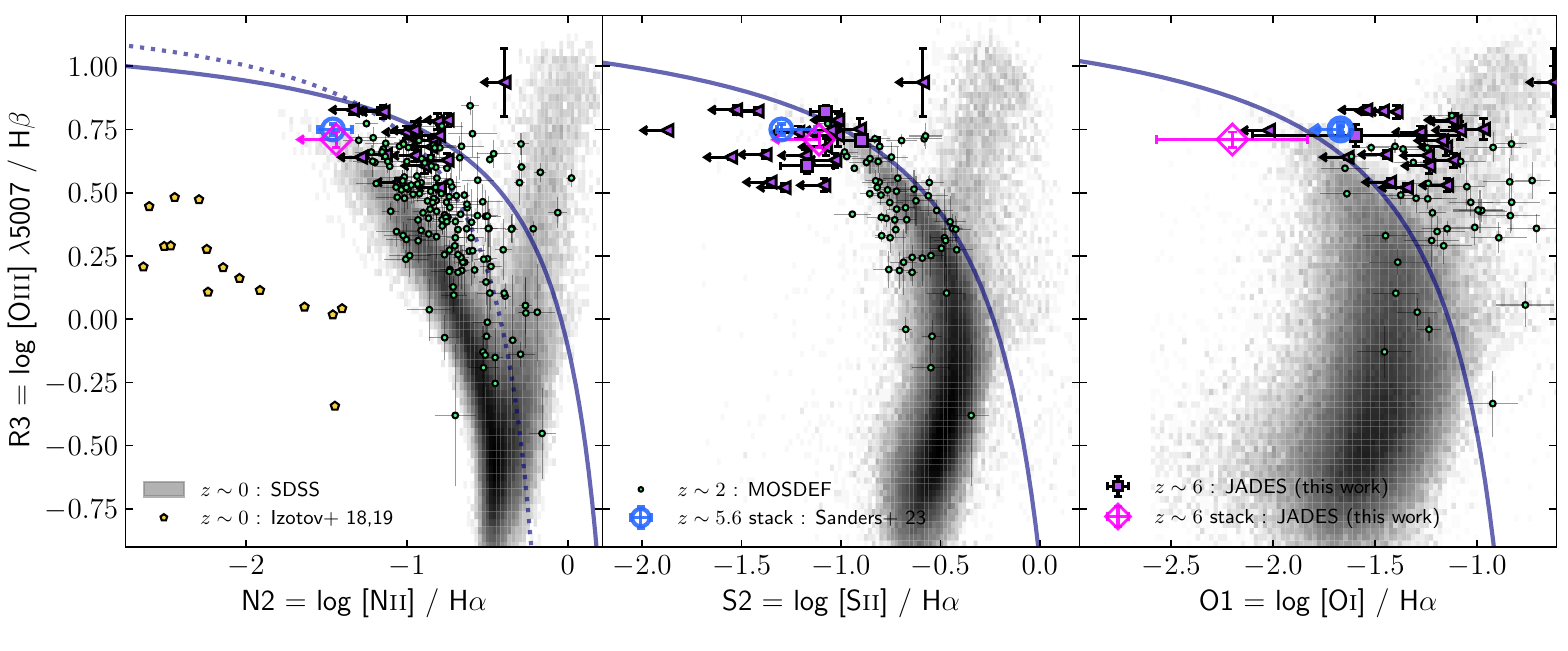}
    \caption{
    BPT and VO87 diagrams showing the comparison between JADES \zsix galaxies and samples across various redshifts.
    Individual measurements are shown as solid purple points, while values dervied from stacked JADES spectra are shown as the open magenta diamonds.
    The background grey 2D PDF shows $z\sim0$ galaxies from SDSS. 
    Extremely metal-poor $z\sim0$ galaxies from \citet{Izotov2018} and \citet{Izotov2019} are shown as yellow pentagons.
    MOSDEF $z\sim2$ galaxies are shown as green circles \citep{Kriek2015}.
    The $z\sim5.6$ composite CEERS spectra presented by \citet{Sanders2023} are shown as the blue open circles.
    The solid navy lines in each panel show the theoretical maximum starbust lines from \citet{Kewley2001}, while the the dotted line in the left panel shows the empirical demarcation derived by \citet{Kauffmann2003}.
    \textit{Left: Classical BPT or `$N2$-BPT' --}  Purple triangles provide 3-$\sigma$ upper limits on the locations of our JADES galaxies since we are unable to detect \NIIl 6583 in any of the 18 galaxies for which we have \grating\, spectral coverage of the \Has complex, although the $R3$ ratio is very well constrained in all cases from PRISM data.
    Even after stacking the grating spectra of all galaxies, we do not recover a robust detection of \NIIl 6583.
    \textit{Centre:} In the $S2$-VO87 diagram (often referred to as `$S2$-BPT'), 3/20 of our JADES galaxies show detections of \SIIs in individual spectra. The tight 3-$\sigma$ upper limits we place on our non-detections highlight that there must be more than an order of magnitude in scatter in the $S2$ ratio within the sample.
    \textit{Right:} We find one tentative detection of \OIl 6300 in our $O1$-VO87 diagram.
    }
    \label{fig:n2_bpt}
\end{figure*}

\subsection{Comparison samples}

In this section, we compare the emission line ratios measured for our JADES \zsix and \zeight samples to various literature samples in several diagnostic diagrams.
To form our baseline $z\sim0$ comparison sample, we use $0.03<z<0.1$ galaxies from SDSS MPA-JHU catalogs\footnote{\url{https://www.sdss3.org/dr10/spectro/galaxy_mpajhu.php}} \citep{Aihara2011} to generate 2D PDFs. We consider only galaxies with spectra flagged as `reliable' in the public release catalog, and select for only galaxies with prominent emission lines by making a signal-to-noise cut of $S/N>40$ on \Ha. For diagnostic diagrams that include the \NeIIIl 3869 line, we additionally apply a cut of $S/N_{\lambda3869}>5$. In the $N2$-BPT, $S2$-VO87, and $O1$-VO87 diagrams we plot all sources that pass these criteria (including any AGN), however for all other plots, we additionally select for only star-forming galaxies using the \citet{Kauffmann2003} criterion.

We consider $z\sim2$ galaxies from the MOSDEF survey public release emission line catalog\footnote{\url{https://mosdef.astro.berkeley.edu/for-scientists/data-releases/}} \citep{Kriek2015, Reddy2015}.
For this sample we impose signal-to-noise cuts of $S/N>5$ on each emission line and plot only detections (i.e. we do not show any limits).
We additionally select only galaxies with a sky line flag with $<0.2$ for all lines considered, as recommended by the catalogue release. For lines that are flagged as being near the edge of the spectrum, we also select only those flagged as having reliable flux.
For diagrams involving $N2$, $S2$ or $O1$, the MOSDEF sample spans from $1.2<z<2.6$ with median $z=2.1$, while for diagrams involving only the bluer lines, it probes slightly higher redshifts: typically $2.0\lesssim z \lesssim 3.6$ with median of $z\approx2.3$.

A number of line ratio measurements have already been reported from NIRSpec at $z\gtrsim6$. The JWST Early Release Observations (ERO; \citealt{Pontoppidan2022}) provided emission line measurements for three galaxies at $z>7.5$, one of which was reported to have an extremely low metallicity of 12+log$(O/H)=6.99\pm0.11$ \citep{Curti2023_ero}.
We use these three galaxies as comparison points in oxygen-based diagrams, adopting the recently re-measured ratio values from \citet{Nakajima2023}.
We compare our individual galaxy measurements with individual $4.5 \leq z \leq 8.0$ galaxies targeted in the GLASS survey, which leverages gravitational lensing \citep{Mascia2023}.
Additionally, we compare to composite spectra presented in \citet{Sanders2023} and \citet{Tang2023} from the Cosmic Evolution Early Release Science (CEERS) survey. 
The $z\sim5.6$ and $z\sim7.5$ redshift bins in \citet{Sanders2023} are well matched in redshift to the sample presented here. Those authors report a median stellar mass of log$(M_*/M_\odot) = 8.57^{+0.04}_{-0.13}$ for their $z\sim5.6$ composite, while they do not estimate masses in their $z\sim7.5$ bin.
\citet{Tang2023} present one composite spectrum of 16 $z\gtrsim7$ galaxies from CEERS with median redshift $z=7.7$. Those authors report a median UV magnitude of $M_{UV}=-20.6$. This is considerably brighter than our median magnitude of $M_{1500}=-18.71$, indicating that the shallower CEERS spectroscopy is probing generally brighter galaxies that our JADES data.

We also consider a number of possible `high-redshift analogs' identified from low-redshift samples by various authors.
`Green pea' galaxies are a population of galaxies at $z\sim0$ with unusually high equivalent-width emission of \OIIIl 5007 ($\sim$1000 \AA{}) \citep{Cardamone2009, Yang2017_GreenPea}. These galaxies are characterised by masses of $M_*\sim10^{8.5-9.5}M_\odot$ and high star-formation rates ($\sim10 M_\odot$ yr$^{-1}$), analogous to galaxies at higher redshifts.
We consider also `blueberry' galaxies, which are a population of $z\sim0$ dwarfs with even more extreme properties than green peas, having masses of $M_*\sim10^{6.5-7.5}M_\odot$, very high ionisation (\OIII/\OIIs $\gtrsim10$) and very low metallicities (12+log$(O/H)\approx7.1 - 7.8$) \citep{Yang2017_Blueberry}. 
We also consider samples of the most metal-poor galaxies known in the low-redshift Universe \citep{Izotov2018, Izotov2019}.

The $z\sim0$ comparison samples are corrected for dust assuming a \citet{CCM1989} law with $R_V=3.1$ where the $A_V$ is derived from the H$\alpha$/H$\beta$ decrement assuming an intrinsic ratio of 2.86.
Meanwhile, the $z\sim2$ MOSDEF sample is corrected in an identical manner to that used for our $z\sim6$ sample, described in Section~\ref{sub:spectral_fitting}.

\begin{figure*}
    \centering
    \includegraphics[width=0.95\textwidth]{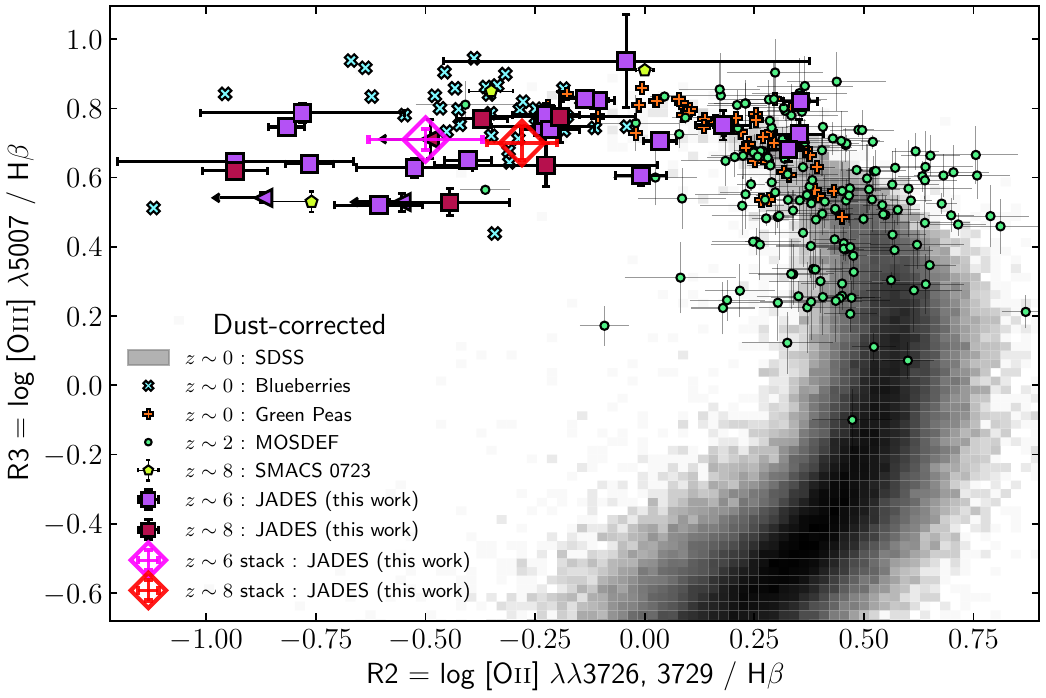}
    \caption{
    The $R2-R3$ diagram showing our two JADES sub-samples at \zsix (purple points) and \zeight (maroon points). Comparison samples from SDSS and MOSDEF are shown as in Figure~\ref{fig:n2_bpt}.
    We additionally show comparison with two populations of $z\sim0$ extreme starbursts: green peas (\citealt{Yang2017_GreenPea}; orange plusses) and blueberries (\citealt{Yang2017_Blueberry}; blue crosses). The three well-studied $z>7.5$ galaxies from the SMACS 0723 ERO observations are shown as lime green pentagons, adopting the ratio values presented by \citet{Nakajima2023}.
    Unlike the BPT and VO87 diagrams, the majority of our sample have $>3\sigma$ detections of all lines in individual spectra, revealing a large diversity within the sample.
    Ratios measured from stacked JADES spectra are shown as the open magenta and red diamonds (see Section~\ref{sub:stacking}).
    }
    \label{fig:r2_r3}
\end{figure*}

\subsection{Excitation properties of $z\sim6$ galaxies: BPT and VO87 diagrams}
\label{sub:n2_bpt}

The left panel of Figure~\ref{fig:n2_bpt} shows our $z\sim6$ JADES sample plotted onto the so-called $N2$-BPT diagram \citep{BPT1981}.
As described in Section~\ref{sub:grating_fitting}, we find no convincing detections of \NIIl 6583 in any of our JADES $z\sim6$ sample (purple points).
Figure~\ref{fig:n2_bpt} also shows comparison samples of $z\sim0$ galaxies from SDSS (background 2D PDF), extremely-metal poor galaxies at $z\sim0$ \citep[][yellow pentagons]{Izotov2018, Izotov2019}, $z\sim2$ galaxies from MOSDEF (green circles), and the $z\sim5.6$ composite CEERS spectrum from \citet{Sanders2023}.

Based on the $R3$ ratios alone, we can already see that these JADES \zsix galaxies span the upper extreme of what is observed for galaxies in SDSS and MOSDEF, with the 18 galaxies shown here having a median $R3$ of 0.74, and all lying above $R3>0.53$. 
Considering the $N2$ ratio, SDSS galaxies lying along the star-forming BPT sequence with $0.72<R3<0.76$, have a median value of $N2_{\rm median}=-1.4$. None of our JADES galaxies have upper limits on the $N2$ ratio which are inconsistent with this value at the 3-$\sigma$ level, while the majority of our sample have individual upper limits of only $\lesssim-0.9$ (see also Table~\ref{tab:line_ratios}).
Beyond our individual non-detections, we find that we are unable to robustly detect \NIIl 6583 even in a stack of all 18 spectra, placing a 3-$\sigma$ upper limit of $N2<-1.44$. 
Previous studies have widely observed that $z\sim2$ galaxies are generally offset from the $z\sim0$ $N2$-BPT sequence, toward higher values of $N2$ at fixed $R3$ (\citealt{Kewley2013_evolution, Steidel2014, Strom2017, Runco2022}; green MOSDEF points in Figure~\ref{fig:n2_bpt}).
Given the anti-correlation between $N2$ and $R3$ in this sequence, and the high $R3$ values measured in our sample, it is not possible to characterise from individual measurements what fraction of the JADES sample would lie significantly `above' the $z\sim0$ sequence. However the upper limit imposed from our stacked spectrum suggests the median value at this redshift might be more aligned with the $z\sim0$ sequence than measurements from $z\sim2$ samples.

\citet{FengwuSun2022b} made emission line measurements at $z\sim6$ with NIRCam slitless spectroscopy, with two modest detections of \NIIl 6583 showing much higher $N2$ values of $N2=-0.77$ and $N2=-0.69$ (see also \citealt{FengwuSun2022a}).
However, those two galaxies were more than an order of magnitude brighter ($m_{\rm F444W} < 25$) than the JADES galaxies probed here ($m_{\rm F444W; median} = 28.1$, for galaxies with NIRCam coverage), suggesting \citet{FengwuSun2022b} are probing more massive systems.
\citet{Sanders2023} present six individual galaxies at $z>5$ with detections of \NIIl 6583 from CEERS spectroscopy, also showing much higher $N2$ values ($N2\gtrsim-1.0$). Taking a composite spectrum of all 38 galaxies in the CEERS sample from $5.0<z<6.5$ they find $N2=-1.46^{+{0.12}}_{-{0.09}}$, quoting a relatively high median stellar mass of log $(M_*/M_\odot)=8.57^{+0.04}_{-0.13}$.
Our inability to detect $N2$ in any of our sample may indicate none of our galaxies have reached the same levels of metal-enrichment as some of the brighter individual galaxies observed in \citet{FengwuSun2022a} or \citet{Sanders2023}.
However, comparison with low-redshift galaxies with extremely low metallicity \citep[][yellow pentagons]{Izotov2018, Izotov2019} highlights that the 7 hour deep \grating\, spectra presented here are well short of being sufficient to probe \NIIs emission from the most metal-poor systems in the early Universe.
We see in the left panel of Figure~\ref{fig:n2_bpt} that, at $z\sim0$, these most metal poor systems often show $N2$ ratios well below $N2<-2$, and have generally milder values of R3 ($0.0 < R3 < 0.5$).
At $z\sim0$, the sequence of $R3$ in galaxies is known to be double-valued with metallicity, increasing from the lowest metallicities up to a turnover around $12+\text{log}(O/H)\approx8$, before decreasing toward higher metallicity \citep[e.g.][]{Curti2020}.
The fact that our JADES $R3$ ratios are notably higher than those from the \citet{Izotov2018} and \citet{Izotov2019} galaxies suggests we are not probing galaxies below this turnover and are more likely sampling galaxies only down to around $\sim0.1 \times Z_\odot$, where the $R3$ sequence is relatively flat \citep{Curti2020}.
Understanding N/O abundance ratios in metal-poor galaxies in the Universe will be critical to developing chemical evolution models of galaxies \citep{Maiolino2019, HaydenPawson2022}. This study demonstrates that probing N/O abundances in the most metal poor galaxies at $z\gtrsim5.5$ with emission lines will likely be very challenging even with very deep spectra.

The centre panel of Figure~\ref{fig:n2_bpt} shows our JADES sample plotted onto the $S2$-VO87 diagram \citep{VO1987}\footnote{This diagram is often referred to as the $S2$-BPT diagram. However, it was first introduced by \cite{VO1987}.}.
Because the \SIIll 6716, 6731 doublet is more widely separated in wavelength from H$\alpha$, we can measure this ratio directly from the much deeper PRISM spectrum\footnote{Note that, although H$\alpha$ would in theory be blended with \NIIs at this resolution, as highlighted above, we see no evidence for the presence of \NIIs in either the PRISM or grating spectra of any galaxies in our sample. From our grating upper limits we can constrain that blending between H$\alpha$ and \NIIs would contribute no more than 6 \% (0.02 dex) to a bias the observed $S2$ ratio.}.
Consequently, we observe three \SIIs detections at greater than 3-$\sigma$. These detections are broadly consistent with lying along the upper extension of the $z\sim0$ sequence and $z\sim2$ sequence.
The deeper PRISM spectra also allow us to place much tighter upper limits on the S2 ratio for the 17/20 non-detections in our sample.
In some cases the 3-$\sigma$ upper limits are more than an order of magnitude lower in $S2$ than our three detections, suggesting there is significant scatter in $S2$ ratios within the $z\sim6$ population.
The composite $z\sim5.6$ CEERS spectrum from \citet{Sanders2023} yields $S2=-1.30^{+0.17}_{-0.02}$, which is broadly in line with the three JADES galaxies with measurable \SII. 
This suggests that stacks based on the shallower, but wider area, CEERS spectroscopy are biased toward galaxies that are more in line with the tip of the $z\sim0$ sequence, which are likely more evolved systems.

In the right panel of Figure~\ref{fig:n2_bpt} we next plot $O1$ = \OIl 6300/\Has against the $R3$ ratio. 
As with the $S2$ ratio, we can again derive \OI/\Has constraints directly from the PRISM spectra.
We find one tentative \OIs detection in our sample (JADES-GS+53.15613-27.77584), indicating the presence of neutral gas in this galaxy. 
The measured $O1$ ratio for this galaxy places it above the theoretical maximum starburst excitation limit from \citet{Kewley2001}, which could suggest the presence of shock-heated gas in the ISM \citep{Sutherland2017}.
Interestingly, this galaxy is not one of the three galaxies with a \SIIs detection from Figure~\ref{fig:n2_bpt}. However, it does have the second lowest $O32$ ratio of galaxies in this sample (Table~\ref{tab:line_ratios}).
Overall, the individual $O1$ constraints presented here are not strongly constraining relative to the \citet{Kewley2001} demarcation line.
The measurement from our stacked spectrum suggests the median $O1$ in this sample could be consistent with the upper extension of the $z\sim0$ star-forming sequence.
However, we showed above that there is significant diversity in $S2$ ratios within our sample. If this diversity also holds for $O1$, it is clear that larger samples probing even deeper would be required to characterise the the sample properties of $O1$ at $z\sim6$.

\begin{figure*}
    \centering
    \includegraphics[width=0.95\textwidth]{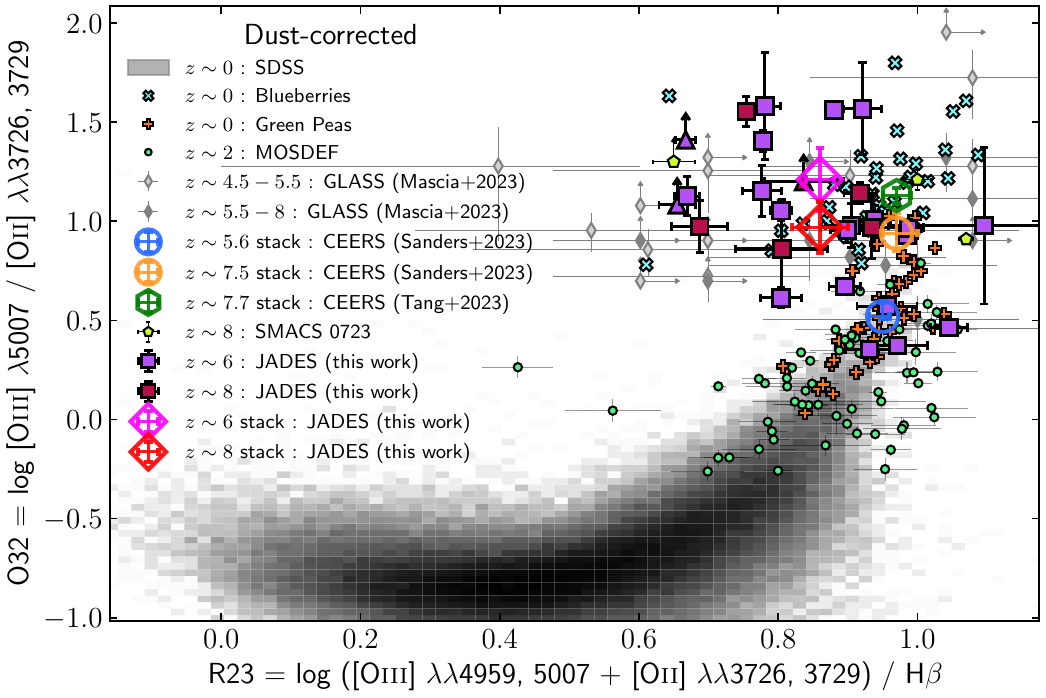}
    \caption{
    The $R23$-$O32$ diagram, probing the ionisation and excitation of the ISM.
    The JADES sub-samples at $z\sim6$ (purple points) and $z\sim8$ (maroon points) lie at generally high excitation an high ionisation.
    Comparison samples, including SDSS, MOSDEF, green peas, blueberries, and SMACS 0723, are shown as in Figure~\ref{fig:r2_r3}.
    $z\sim4.5-8$ galaxies from GLASS are shown as grey diamond \citep{Mascia2023}.
    Blue and orange open circles show the $z\sim5.6$ and $z\sim7.5$ composite CEERS spectra from \citet{Sanders2023}, while the $z\sim7.7$ composite CEERS spectrum from \citep{Tang2023} is shown as the green open hexagon.
    Stacked JADES spectra are shown as the open magenta and red diamonds.
    }
    \label{fig:r23_o32}
\end{figure*}

\begin{figure*}
    \centering
    \includegraphics[width=0.95\textwidth]{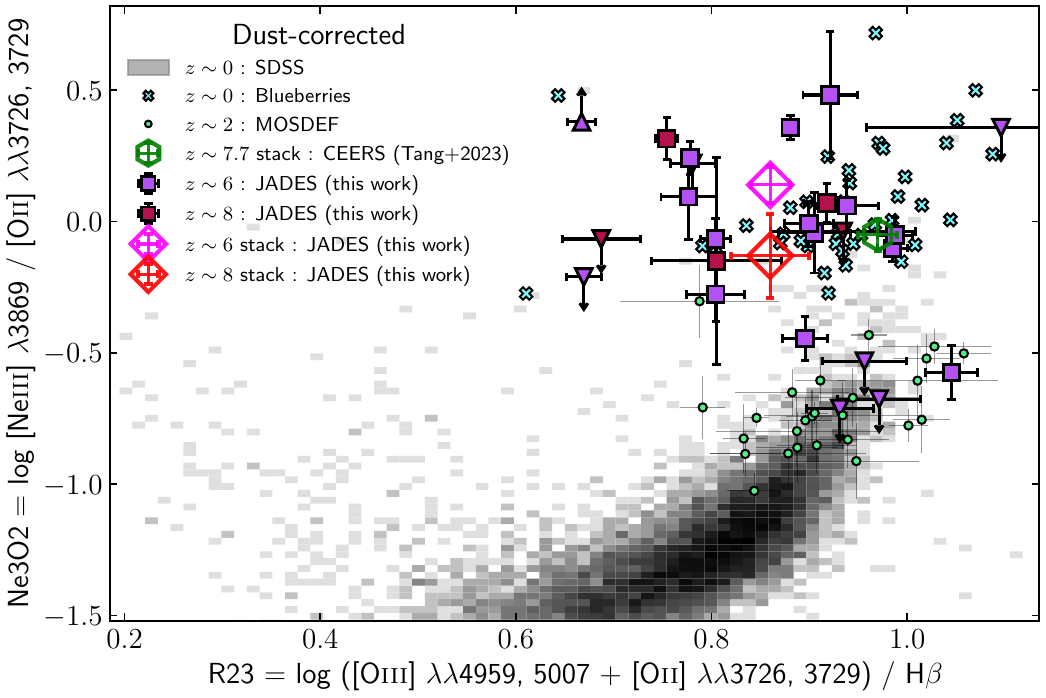}
    \caption{
    The $R23$-$Ne3O2$ diagram, probing the ionisation and excitation of the ISM. $R23$-$Ne3O2$ closely resembles the $R23$-$O32$ diagram with the additional advantage that the \NeIIIl 3869 / \OIIll 3726,329 ratio is less sensitive to the applied dust-correction, or any wavelength-dependent flux calibrations.
    JADES galaxies are plotted as in Figure~\ref{fig:r23_o32}.
    Fewer literature samples present \NeIIIs ratios so we show only SDSS, blueberries, MOSDEF, and the $z\sim7.7$ composite CEERS spectrum from \citep{Tang2023}. These are each plotted in the same way as in previous figures.
    }
    \label{fig:r23_ne3o2}
\end{figure*}

\subsection{Excitation properties from $z\sim5.5-9.5$: $R2-R3$ diagram}

In Figure~\ref{fig:r2_r3} we instead consider the $R2$ = \OIIll 3726, 3729 / H$\beta$ ratio. The longer wavelength baseline of this ratio has the drawback of being much more sensitive to the reddening correction applied compared to the $N2$-BPT, $S2$-VO87, and $O1$-VO87 diagrams discussed in Section~\ref{sub:n2_bpt}.
However, the shorter rest-frame wavelength of these emission lines means that the $R2-R3$ diagram can be used to explore the ISM conditions of galaxies up to $z\sim9.5$ with \emph{JWST}/NIRSpec.
\OIIs is typically intrinsically much brighter than \NII, \SII, or \OI. Moreover, like \SIIs and \OI, ratio measurements can be taken directly from the PRISM spectrum.
As a result, not only does Figure~\ref{fig:r2_r3} include the additional six galaxies from the $z\sim8$ sample, we also have a much higher detection rate, with 24/27 galaxies in our combined sample having detections of \OII.

We see immediately in Figure~\ref{fig:r2_r3} that there is more than an order of magnitude in scatter in the $R2$ ratio, suggesting there is significant diversity in the ISM conditions of galaxies within this sample.
We find a median value of $R2=-0.23$ with standard deviation of 0.38 among our 24 detections, significantly offset from the median value from MOSDEF of $R2=0.38$. Instead, we find that the range of $R2$ values observed for this sample are much more aligned with those observed in extreme $z\sim0$ dwarf starbursts such as green peas (\citealt{Cardamone2009, Yang2017_GreenPea}; orange plusses) and blueberries (\citealt{Yang2017_Blueberry}; blue crosses).

Both $R2$ and $R3$ are known to form a double-valued sequence in metallicity \citep[e.g.][]{Curti2020}. In $z\sim0$ galaxies, $R3$ does not change strongly with metallicity between $\sim0.1-0.5\times Z_\odot$. On the other hand, $R2$ turns over at higher metallicities between $\sim0.5-0.8\times Z_\odot$, at a maximum value of $R2\approx0.5$ \citep{Curti2020}. 
The combination of highly consistent $R3$ ratios but wildly varying $R2$ ratios indcates that our sample likely falls into the metallicity range of $\sim0.1-0.5\times Z_\odot$.
We have five galaxies in our JADES sample with $R2>0.0$, overlapping with the region inhabited by typical $z\sim2$ MOSDEF galaxies. These may represent systems which are more evolved and more metal-rich.
Below its metallicity turnover, $R2$ values in the $z\sim0$ sample from \citet{Curti2020} drop steeply with decreasing metallicity. 
The lowest $R2$ values measured in our sample lie below $R2<-0.75$, below the range probed by the \citet{Curti2020} calibration.
However we note that the low $R2$ portion of our sample exhibits a fair resemblence to blueberry galaxies from \citet{Yang2017_Blueberry} which have been show to have very low-metallicties $\lesssim0.1\times Z_\odot$.
However, given that $R2$ is strongly affected by the ionisation parameter, inferring metallicity values from this ratio is likely not robust.
Nonetheless, there is clearly a large range of metallicity being probed in this sample.

We see no evidence for evolution of the $R2$ ratio between our $z\sim8$ and $z\sim6$ samples, albeit with only six $z\sim8$ galaxies.
We note that all of our $R2>0$ galaxies are from the $z\sim6$ sample; however, the measurements from stacked spectra in the two samples actually yield a higher value for the $z\sim8$ sample in Figure~\ref{fig:r2_r3} (open diamonds), likely affected by uncertainty due to small sample size.
This suggests that there is no rapid time evolution in this diagram at this epoch and that the scatter is instead driven by sample diversity.

\subsection{Ionisation-excitation diagrams: R23-O32 \& R23-Ne3O2}
\label{sub:r23_o32}

Figure~\ref{fig:r23_o32} shows the $R23$-$O32$ diagram with our JADES sample plotted against the same literature comparison samples shown in Figure~\ref{fig:r2_r3}.
$R23$ is often taken as an indication of total excitation, as it encompasses emission lines from both singly and doubly ionised oxygen, while the $O32$ ratio is sensitive to ionisation parameter \citep[e.g.][]{Kewley2019}.

It is in this diagram that the offset of $z\gtrsim6$ from typical $z\sim0$ and $z\sim2$ galaxies is perhaps most striking: $O32$ ratios are clearly much higher in JADES $z\sim5.5-9.5$ galaxies compared to MOSDEF galaxies at $z\sim2$.
Across our combined sample we find a median $O32$ value of 0.98 with a standard deviation of 0.36, with 18 of our galaxies have robust detections showing $O32>0.75$. This is much higher than MOSDEF galaxies plotted in Figure~\ref{fig:r23_o32} which have a median $O32$ value 0.18 and are already tracing the upper end of the $z\sim0$ relation.
This indicates that galaxies across our sample exhibit very high ionisation parameters, much larger than what are seen in typical galaxies at $z\sim0-3$.
However, we can see in Figure~\ref{fig:r23_o32} that our JADES sample largely overlaps with the region spanned by blueberries \citep{Yang2017_Blueberry} and green peas \citep{Yang2017_GreenPea}, which are examples of $z\sim0$ galaxies with extreme star-formation activity. Comparable $O32$ ratios have also been observed in $z\sim2-4$ Lyman-$\alpha$ emitters \citep{Nakajima2016, Tang2021} and $z\sim1.3-2.4$ extreme \OIIIs emitters \citep{Tang2019}.
It is therefore likely that the high $O32$ ratios observed in our sample are indicative of intense star formation activity.

We find that our $z\sim6$ sample is significantly offset from the $z\sim5.6$ stack presented in \citet{Sanders2023}, with our composite $O32$ value found to be more than a factor of four larger. Indeed where individual detections from CEERS are reported in \citet{Sanders2023} they typically have $O32<0.75$.
Mirroring what was observed in our $R2$ sequence from Figure~\ref{fig:r2_r3}, we find six galaxies with much lower $O32$, more in line with the composite spectrum from \citet{Sanders2023}.
Given the resemblance between the upper end of our $O32$ sample and dwarf starbursts such as blueberries, it is likely that this portion of our sample is probing higher star-formation rate and lower-metallicity galaxies (analogous to blueberries), while the handful of lower $O32$ galaxies presented here in alignment with the \citet{Sanders2023} $z\sim5.6$ composite could be probing more evolved systems.

On the other hand, the composite CEERS values at $z\sim7.5$ and $z\sim7.7$, from \citet{Sanders2023} and \citet{Tang2023} respectively, are in closer agreement with our composite spectra, albeit still offset toward slightly higher $R23$.
This could suggest that the fraction of more evolved, metal-enriched systems is increasing rapidly from $z\sim8$ to $z\sim6$, despite the bulk of our UV-selected $z\sim6$ sample showing no significant evolution from our $z\sim8$ sample.
Addressing this question in more detail requires larger samples of deep spectroscopy where properties of individual galaxies can be isolated.

By leveraging gravitational lensing, the measurements of $4.5\lesssim z \lesssim 8$ galaxies from \citet{Mascia2023} can more readily probe fainter and lower mass systems. These are similarly offset from the \citet{Sanders2023} $z\sim5.6$ composite and largely overlap with our JADES sample, spanning a large range in $R23$. Two galaxies are reported by \citet{Mascia2023} as having $R23\lesssim0.6$ despite having high $O32$ comparable with our JADES sample. This indicates that gravitational lensing may be important for study low metallicity galaxies (below the $R23$ turnover) and not simply ultra-deep observations like those presented here.

We note that, as with Figure~\ref{fig:r2_r3}, this diagram relies on ratios with long baselines in wavelength and that dust reddening can have a significant effect. 
The dust-correction (described in Section~\ref{sub:line_ratios_measure}) is most significant on the $O32$ ratio here, with larger corrections serving to reduce $O32$ ratios.
From the corrections we apply, all but two galaxies have corrections that shift $O32$ by less than 0.15. The offset between the \citet{Sanders2023} $z\sim5.6$ composite and our $z\sim6$ stack cannot be explained by the dust correction; `correcting' our value down to $O32\approx0.5$ would require $E(B-V)\gtrsim1$, significantly larger than our measured value of $E(B-V)=0.06$.

Nonetheless, in Figure~\ref{fig:r23_ne3o2} we show our JADES $z\sim6$ and $z\sim8$ samples with $Ne3O2$ replacing $O32$ to provide the ionisation axis, which provides the bonus that $Ne3O2$ is essentially unaffected by dust extinction and the associated correction uncertainties\footnote{Note that at the spectral resolution of our PRISM observations, \NeIIIl 3869 is partially blended with the He {\sc i} $\lambda$3889 line. See discussion in Section~\ref{sub:spectral_fitting}.}.
Despite the lower detection rate, and generally much larger measurement uncertainties associated with switching from \OIIIl 5007 to the intrinsically much fainter \NeIIIl 3869, we see the same story reflected here with the JADES sample exhibiting large scatter (median $Ne3O2=-0.11$ with standard deviation 0.29) and generally offset from SDSS and MOSDEF galaxies, instead being more aligned with blueberry dwarf starbursts.
This again supports the conclusion that our JADES sample is probing a diverse sample of galaxies, many of which exhibit very high ionisation parameters analogous to those observed in local compact starbursts.
\citet{Tang2023} presented $Ne3O2$ measurements from their $z\sim7.7$ composite CEERS spectrum. These are marginally lower than our $Ne3O2$ composite values. As in Figure~\ref{fig:r23_o32}, the larger offset is again the 0.1 dex difference in $R23$.
This offset could reflect an increase in metallicity from our sample to the \citet{Tang2023} galaxies which are reported has having typically brighter UV magnitudes (median $M_{\rm UV}=-20.6$ from \citealt{Tang2023}).

Finally, we note that \NeIIIl 3869 emission may become important for confirming spectroscopic redshifts at $z\gtrsim10$ where \OIIIs $\lambda\lambda$ 4959, 5007 is no longer observable within the wavelength coverage of NIRSpec.
We find the \NeIIIl 3869 emission to be brighter than the \OIIs doublet in at least 8/27 galaxies in our sample. Given that these cases should represent the more extreme examples (highest ionisation parameters and lowest metallicities), \NeIIIl 3869 may be one of the most accessible emission lines in searches for the highest redshift galaxies. However, existing deep spectroscopic observations of such galaxies have already demonstrated that observing any emission lines at those redshifts can be challenging \citep{CurtisLake_z10, Robertson_z10}.

\section{Discussion} \label{sec:discussion}

\subsection{ISM conditions in the Epoch of Reionisation}
\label{sub:model_discussion}

\begin{figure*}
    \centering
    \includegraphics[width=\textwidth]{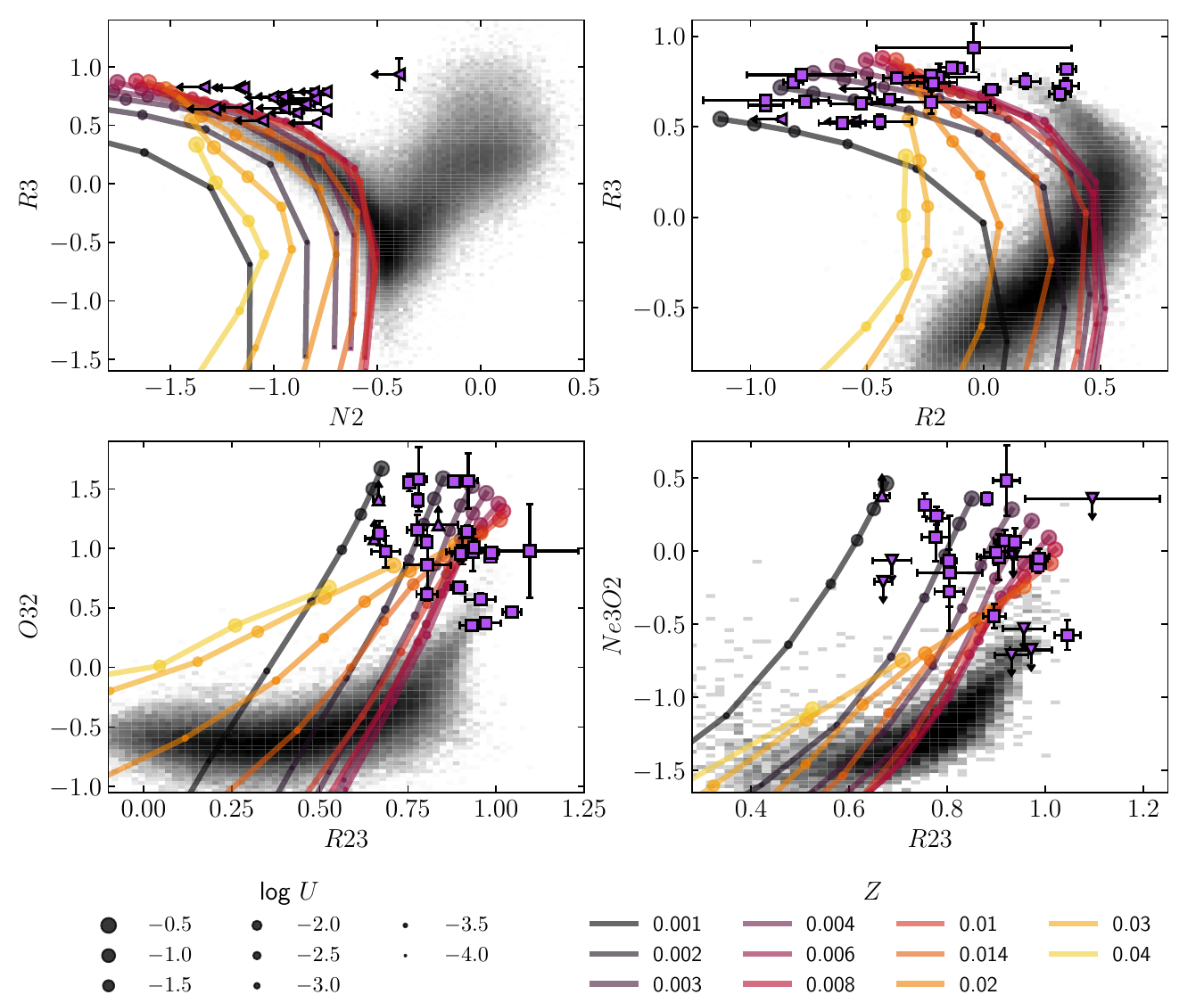}
    \caption{
    JADES $z\sim5.5-9.5$ galaxies compared to photoionisation models. 
    Measured ratios are shown as purple points, as in Figures~\ref{fig:n2_bpt}~-~\ref{fig:r23_ne3o2} except that all galaxies are plotted in the same colour, regardless of redshift. SDSS galaxies are shown in the same manner as Figures~\ref{fig:n2_bpt}~-~\ref{fig:r23_ne3o2}.
    Each line shows a set of models with constant metallicity with values of $Z=$ [1, 2, 3, 4, 6, 8, 10, 14, 20, 30, 40] $\times10^{-3}$ increasing from black to yellow. Assuming $Z_\odot=0.0134$ \citep{Asplund2009}, these correspond to $Z/Z_\odot=$ 0.07, 0.15, 0.22, 0.30, 0.45, 0.60, 0.75, 1.0, 1.5, 2.2, 3.0.
    Ionisation parameter varies along the grid line, increasing with marker size, with values of log $U=$ $-4.0$, $-3.5$, $-3.0$, $-2.5$, $-2.0$, $-1.5$, $-1.0$, $-0.5$. Details of the photoionisation models are described in Section~\ref{sub:model_discussion}.
    }
    \label{fig:grids}
\end{figure*}

Section~\ref{sec:results} showed that our $z\sim5.5-9.5$ sample from JADES exhibits significant diversity in ISM conditions.
The measured ratios highlight a clear difference between this high-redshift sample and typical galaxies at $z\sim0-3$, although parallels can be drawn instead with extreme starburst populations of low-redshift galaxies, such as blueberries and green peas. In this section we use photoionisation models to explore the parameter space spanned by our sample.

Metallicity and ionisation parameter are the dominant parameters driving the emission line ratios of H{\sc ii} regions \citep{Kewley2019}.
In Figure~\ref{fig:grids}, we compare our measured ratios to photoionisation model grids in four of the diagnostic plots highlighted in Section~\ref{sec:results}.
These models are calculated with {\sc CLOUDY} v17 \citep{Ferland2017_CLOUDY} adopting input ionising spectra from BPASS v2.0 \citep{Eldridge2016_BPASSv2, Stanway2016_BPASSv2}, assuming an instantaneous burst with an age of 3 Myr.
The models are plane-parallel and assume a 1 kpc thick shell with $n_H=100$ cm$^{-3}$.
The inner radius is set to $10^{30}$ cm and ionisation parameter is varied between $-4.0<$ log $U<-0.5$ in steps of $0.5$. Models are calculated at 11 values of metallicity: $Z=$ [1, 2, 3, 4, 6, 8, 10, 14, 20, 30, 40] $\times10^{-3}$ (or $Z/Z_\odot=$ 0.07, 0.15, 0.22, 0.30, 0.45, 0.60, 0.75, 1.0, 1.5, 2.2, 3.0, assuming $Z_\odot=0.0134$; \citealt{Asplund2009}). Metal abundances ratios are assumed to follow solar abundance patterns \citep{Asplund2009}, and the depletion of metals onto dust is accounted for.

Figure~\ref{fig:grids} highlights that photoionisation models driven by SEDs of young stars can reproduce the parameter space populated by our JADES galaxies with only fairly standard assumptions.
This indicates we do not need to invoke alternative heating sources (such as AGN or shocks) to explain our findings.
We see from this simple comparison that the large scatter observed in our emission line ratio sample likely reflects a diversity in metallicity and ionisation parameter among these galaxies.
Our most extreme galaxies (i.e. highest $O32$) are only reproduced by the models with very low-metallicity ($0.07-0.15 \times Z_\odot$) and very high ionisation parameters ($-2.0 \lesssim$ log $U \lesssim -1.0$).
These ionisation parameters are signifcantly higher than those typically observed in $z\sim0$ \HII\, regions \citep{Dopita2000} or $z\sim2-3$ galaxies \citep{Sanders2016}. 
We note that we have not explored the effect of harder radiation fields, which would increase $O32$ at a fixed ionisation parameter \citep[e.g.][]{Sanders2016}.
Indeed, many studies at $z\sim2-3$ have argued that harder radiation fields, driven by super-solar $\alpha$/Fe abundance ratios, are a dominant effect in driving the $z\sim2$ $N2$-BPT offset \citep{Steidel2016, Strom2017, Topping2020a, Topping2020b, Cullen2021}.
Given the long timescale of iron enrichment via Type Ia supernovae, one would of course expect that these $z\gtrsim5.5$ galaxies would be $\alpha$-enhanced, and models with non-solar abundance ratios would be more appropriate for parameter estimation.

We can see from these comparisons that our sample likely spans quite a range in metallicity. At lower $O32$ values, we see a turnover in the models where metallicity and ionisation parameter become highly degenerate. However, model metallicities of at least $0.45 \times Z_\odot$ are required to approach our highest $R2$ and $R23$ values.
The $R2$, $R3$, and $R23$ ratios are all known to exhibit double-valued relations \citep[e.g.][]{Curti2020} in metallicity. We tentatively observe that our highest $O32$ galaxies tend to have the the lowest $R23$, which reverses the trend seen in SDSS galaxies (grey 2D PDF), suggesting we are sampling the turnover of this relation.
Some of our `mildest' galaxies (i.e. highest $R2$, lowest $O32$) have $R2\approx0.4$ and $R23\approx1.0$ which are broadly consistent with the turnover observed for these ratios in $z\sim0$ galaxies from \citet{Curti2020}. This suggests these galaxies are the most chemically evolved systems in our sample and may have metallicities as high as $0.3-0.5\times Z_\odot$.
Indeed, these are also the points that are in best agreement with the \citet{Sanders2023} $z\sim5.6$ composite spectrum. This suggests the shallower CEERS spectroscopy is probing on average more chemically evolved systems than those targeted in JADES.

Overall, it is clear from this simple model comparison that a range of ISM conditions is required to explain the diversity of emission line ratios observed in this sample.
As noted in Section~\ref{sec:results}, our sample largely spans the same regions of emission-line-ratio-space inhabited by low-redshift extreme starbursts such as blueberries and green peas.
Thus, the emerging picture is that the galaxies in our sample are likely low-metallicity ($\sim0.1-0.3 \times Z_\odot$) systems undergoing intense periods of star-formation driving strong radiation fields with high ionisation parameters ($-2.0 \lesssim$ log $U \lesssim -1.0$) in the ISM.

\subsection{High O32: Implications for escape fraction and reionisation}

The high $O32$ ratios presented in Figure~\ref{fig:r23_o32} could have implications for interpreting the ionising photon escape fraction of these $z\sim5.5-9.5$ galaxies.
When and how the Universe was reionised remains one of the most important open questions in high-redshift astrophysics.
The efficiency with which ionising photons are produced within galaxies and the fraction of these which escape ($f_{\rm esc}$) are two important quantities required to understand the contribution of galaxies to this process.
Unfortunately, $f_{\rm esc}$ can never be directly measured within the epoch of reionisation itself because any light escaping the galaxy blue-ward of the Lyman-limit will interact with intervening neutral hydrogen.
Thus indirect probes of $f_{\rm esc}$ are highly sought after.

High $O32$ has been proposed as being an indicator of higher $f_{\rm esc}$ \citep[e.g.][]{Nakajima2014, Faisst2016}.
The reasoning is that the high $O32$ ratio selects for highly ionised systems which have a higher likelihood of having density-bounded channels through which ionising photons can escape.
However, some studies have shown that $O32$ does not necessarily correlate well with $f_{\rm esc}$; instead finding that results are very dependent on the observed line of sight, and that high $O32$ can be produced in systems without Lyman continuum (LyC) escape \citep{Paalvast2018, Bassett2019, Katz2020_LyC_leakers}.
Nonetheless, samples of Lyman continuum (LyC) leaking galaxies at low-redshift generally show that the fraction of galaxies with high $f_{\rm esc}$ increases toward higher $O32$, even if the correlation is not tight \citep[e.g.][]{Izotov2016, Flury2022}.
Indeed, the limited number of known LyC leakers at higher redshift typically have high $O32$ ratios \citep{Vanzella2016, Vanzella2020, Vanzella2022, Fletcher2019, Saxena2022}.
Extreme \OIIIs emitters at $z\sim2-3$ have also been shown to exhibit high $O32$ ratios (\OIII/\OIIs $\sim10$), while also preferentially selecting for high equivalent-width Lyman-$\alpha$ emission \citep{Tang2019, Tang2021}.
This evidence therefore seems to suggest that high $O32$ is a necessary but not sufficient condition for high $f_{\rm esc}$.

We have shown in Section~\ref{sub:r23_o32} that our JADES $z\sim5.5-9.5$ galaxies have characteristically high $O32$ ratios, with 21/27 of the sample having \OIII/\OIIs $>5$ and a median value of 9.4. 
Owing to the large scatter in $O32$-$f_{\rm esc}$ relations, we do not attempt to derive constraints on the $f_{\rm esc}$ of these galaxies.
However, the high fraction of our sample that meets the criterion of high $O32$ suggests that the fraction of potential LyC leakers might be increasing to high redshift.

\section{Conclusions} \label{sec:conclusion}

We have presented rest-frame optical emission line ratio measurements in 27 individual galaxies at $5.5 < z < 9.5$ spanning over $-17.0 < M_{1500} < -20.4$, observed as part of the `Deep' tier of the JADES program, which represent the deepest integrations yet taken with \emph{JWST}/NIRSpec.
We find that galaxies in this sample occupy regions of line-ratio space that are offset from those inhabited by `typical' galaxies at $z\sim0$ or $z\sim2$, although generally aligned with more extreme low-redshift populations such as `blueberry' and `green pea' dwarf starbursts \citep{Yang2017_GreenPea, Yang2017_Blueberry}.

The high signal-to-noise achieved in individual spectra from our 28 hour deep PRISM/CLEAR integrations reveal significant intrinsic scatter in the observed line ratios, with our JADES $z\sim5.5-9.5$ galaxies spanning more than an order of magnitude in \OII/H$\beta$ and \OIII/\OIIs ratios.
This highlights the diversity in ISM conditions of galaxies at this epoch.

Despite the depth of our observations, we find that low ionisation lines such as \NIIl 6583, \SIIll 6716, 6731 and \OIl 6300 are very challenging to detect in these systems.
We find no convincing detections of \NIIl 6583 in 7 hour deep \grating\, spectra for any of the 18 galaxies in which we have adequate spectral coverage. Stacking these 18 spectra is still unable to deliver a nitrogen detection. However, the 3-$\sigma$ upper limit suggests our sample median is not offset from the $z\sim0$ $N2$-BPT sequence contrary to what has been observed in $z\sim2$ samples \citep[e.g.][]{Steidel2014}.
In our PRISM spectra we find three modest detections of \SIIll 6716, 6731, while the tight upper limits we derive on other targets suggests more than an order of magnitude in intrinsic scatter in the $S2$ ratio within the sample. The fainter \OIl 6300 line yields only one tentative detection.

The bluer \OIIll 3726, 3729 and \NeIIIl 3869 are much more readily detected throughout the sample. Measured $R2$, $O32$ and $Ne3O2$ ratios particularly highlight the diversity in this sample; we find median values and standard deviations of $R2=-0.23\pm0.38$, $O32=0.98\pm0.36$, and $Ne3O2=-0.03\pm0.27$.
We find that these ratios are generally consistent with very high ionisation parameters up to log $U\sim-1.0$. The sample appears to span a significant range in metallicity of $\sim0.1 - 0.3 \times Z_\odot$, depending on metallicity calibrator assumptions.
These ratios are found to be generally consistent with extreme low-redshift populations, particularly green pea and blueberry dwarf starburst samples.
This is consistent with the suggestion that these $z\sim5.5-9.5$ galaxies are low-mass, low-metallicity galaxies undergoing periods of rapid star formation, driving strong radiation fields.

We compare our sample to composite spectra at $z\sim5.6$ and $z\sim7.5$ from CEERS \citep{Sanders2023, Tang2023}.
We find significant offset between our $z\sim6$ sample median and the $z\sim5.6$ CEERS composite. We instead observe good agreement between this $z\sim5.6$ CEERS composite measurement and the low $O32$ end of our sample. Assuming these galaxies represent the more evolved, metal-enriched extent of our sample, it is clear that the shallower CEERS spectra are preferentially picking up on the more evolved portion of the $z\sim6$ population.
We find no evidence for significant evolution between our $z\sim6$ and $z\sim8$ samples (albeit with a very small sample).
Both our $z\sim6$ and $z\sim8$ sample medians are found to be in somewhat closer agreement with the CEERS $z\sim7.5$ composite spectra, which showed large offset from the CEERS $z\sim5.6$ composite spectrum.
This could indicate that the fraction of moderately metal-enriched systems increases rapidly from $z\sim8$ to $z\sim5.5$, even if our sample median does not show evidence of significant redshift evolution.

Within its first six months of observations, \emph{JWST}/NIRSpec has already shed considerable light on the ISM conditions of $z\gtrsim5.5$ galaxies.
The diversity of ISM conditions presented in this work highlights the need for observations that span the full range of galaxy properties.
This will require a combination of deep observations reaching down to even fainter brightness limits than those presented here, as well as wider area surveys probing rarer, more evolved systems at early epochs.
Characterising the ISM conditions in larger samples of galaxies across this broader parameter space will help chart galaxy assembly across the first $\sim$1 Gyr of cosmic time.

\section*{Acknowledgements}
AJC, AS, AJB, JC \& GCJ acknowledge funding from the ``FirstGalaxies'' Advanced Grant from the European Research Council (ERC) under the European Union’s Horizon 2020 research and innovation programme (Grant agreement No. 789056).
FDE, RM, MC, TJL, JW, LS \& JS acknowledge support by the Science and Technology Facilities Council (STFC), ERC Advanced Grant 695671 "QUENCH".
RM also acknowledges funding from a research professorship from the Royal Society.
JW also acknowledges support from the Fondation MERAC.
SC acknowledges support by European Union’s HE ERC Starting Grant No. 101040227 - WINGS.
ECL acknowledges support of an STFC Webb Fellowship (ST/W001438/1).
The Cosmic Dawn Center (DAWN) is funded by the Danish National Research Foundation under grant no.140.
SA \& BRDP acknowledges support from the research project PID2021-127718NB-I00 of the Spanish Ministry of Science and Innovation/State Agency of Research (MICIN/AEI).
RS acknowledges support from a STFC Ernest Rutherford Fellowship (ST/S004831/1).
H{\"U} gratefully acknowledges support by the Isaac Newton Trust and by the Kavli Foundation through a Newton-Kavli Junior Fellowship.
DJE is supported as a Simons Investigator and by JWST/NIRCam contract to the University of Arizona, NAS5-02015.
EE, MR, BJD, BER \& FS acknowledge support from the NIRCam Science Team contract to the University of Arizona, NAS5-02015.
The work of CCW is supported by NOIRLab, which is managed by the Association of Universities for Research in Astronomy (AURA) under a cooperative agreement with the National Science Foundation.
RB acknowledges support from an STFC Ernest Rutherford Fellowship [grant number ST/T003596/1].
This research is supported in part by the Australian Research Council Centre of Excellence for All Sky Astrophysics in 3 Dimensions (ASTRO 3D), through project number CE170100013.



\bibliographystyle{aa} 
\bibliography{paper_bib}



\appendix

\section{Emission line ratio measurements from stacked spectra}

In Section~\ref{sub:stacking} we described our procedure for measuring emission line ratios from stacked spectra. Here we briefly summarise the values obtained from this analysis.
Table~\ref{tab:stack_properties} outlines the number of galaxies included in each stack and median properties of these samples.
Table~\ref{tab:stack_values} provides the dust-corrected emission line ratios obtained from the resultant stacked spectra. These values are shown as open diamonds in Figures~\ref{fig:n2_bpt}~-~\ref{fig:r23_ne3o2}.

\begin{table}[h!]
\centering
\begin{tabular}{ccccccccccccc}
\hline
Sample & Disperser & $N_{\rm galaxies}$ & $z^\dag$ & $M_{1500}^\dag$ \\
\hline
JADES $z\sim6$ & PRISM/CLEAR & 21 & 5.943 & -18.71 \\
JADES $z\sim6$ & \grating & 18 & 5.945 & -18.73 \\
JADES $z\sim8$ & PRISM/CLEAR & 6 & 7.629 & -18.82 \\
\hline
\end{tabular}
\caption{
Properties of samples underpinning stacked spectra. \\
$^\dag$ Median value of all galaxies included in the stack.
}
\label{tab:stack_properties}
\end{table}

\begin{table}[h!]
\centering
\begin{tabular}{lcccccccccccc}
\hline
Ratio & JADES $z\sim6$ & JADES $z\sim8$ \\
\hline
$N2$ & $<-1.44$ & ... \\
$S2$ & $<-1.11$ & ... \\
$O1$ & $-2.20\pm0.37$ & ... \\
$R3$ & $0.71\pm0.03$ & $0.70\pm0.05$ \\
$R2$ & $-0.50\pm0.13$ & $-0.28\pm0.08$ \\
$R23$ & $0.86\pm0.02$ & $0.86\pm0.04$ \\
$O32$ & $1.21\pm0.16$ & $0.97\pm0.13$ \\
$Ne3O2$ & $0.14\pm0.07$ & $-0.13\pm0.16$ \\
\hline
\end{tabular}
\caption{
Emission line ratio measurements from our stacked JADES spectra. Ratio names are defined in Equation~\ref{eqn:ratios}. Non-detections are quoted as 3-$\sigma$ upper limits. Ratios in this table are derived from stacked PRISM/CLEAR spectra in all cases, with the exception of $N2$ which is derived from stacked \grating\, spectra.
}
\label{tab:stack_values}
\end{table}

\section{Verification of flux calibration}
\label{sec:flux_cal_verify}

\begin{figure}
    \centering
    \includegraphics[width=0.97\columnwidth]{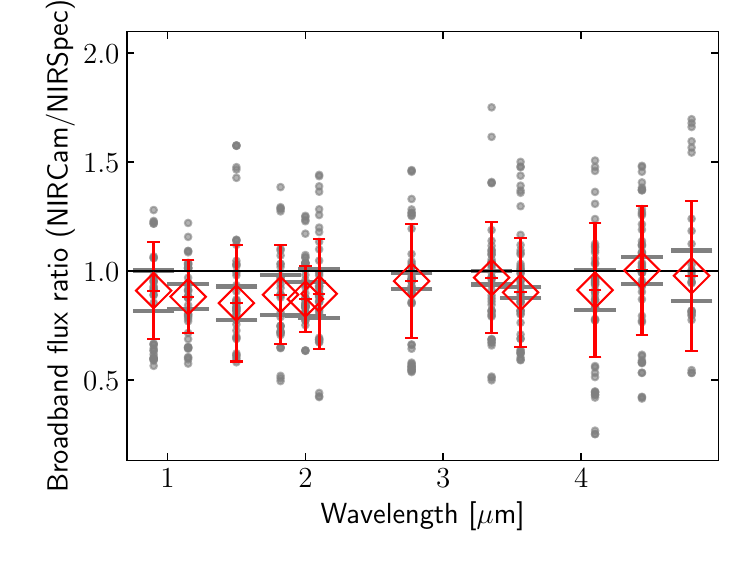}
    \caption{Ratio of apodized broadband fluxes extracted from the projected MSA shutter regions on the NIRCam imaging compared to fluxes obtained from convolving PRISM/CLEAR spectra with the NIRCam filer transmission curves (see text for details). Grey points show values derived for each independent pointing of each galaxy targeted in this work. Horizontal grey marks indicate the typical 1-$\sigma$ measurement uncertainty on the individual measurements in each filter (which encompasses uncertainties from both NIRCam and NIRSpec). Red diamonds and errorbars show the median and standard deviation for each filter.}
    \label{fig:flux_NS_NC_compare}
\end{figure}

As a verification of our spectral flux calibration, we compared the fluxes obtained from NIRCam photometry with those obtained from the PRISM/CLEAR spectra.
Using imaging data from the JADES program \citep{Eisenstein2023, Rieke2023}, we extract apodized fluxes for each filter by extracting the flux within the projected shutter position of each target.
Assuming the MSA pointing uncertainty is negligible, this should represent the total amount of flux in each spectral region incident on the shutter area.
However, the PSFs of NIRCam and NIRSpec are different, and in reality the final flux recorded by NIRSpec is additionally reduced by losses due to diffraction caused by the shutter itself.
Our data reduction pipeline includes a step which corrects for these slitlosses, meaning that the output spectrum should be scaled to the same flux density as the total magnitude of the source (correcting for position within the slit and flux falling outside the slit). As a further check on the relative flux calibration between NIRCam and NIRSpec, we also consider the flux from the NIRCam imaging which would fall within the NIRSpec microshutter aperture. To make a fair comparison, we first remove the slitloss correction applied by the pipeline to the NIRSpec spectrum (to consider just the uncorrected flux within the shutter) and convolve the resultant PRISM/CLEAR spectrum with the NIRCam filter transmission curves.

Figure~\ref{fig:flux_NS_NC_compare} shows the comparison between the NIRCam and NIRSpec fluxes derived in this way, with grey points representing each individual pointing\footnote{Note each target was observed in up to three independent MSA configurations (see Section~\ref{sec:data}), and that these in general have similar, but not identical, shutter projections.} of each target galaxy. Typical statistical uncertainties for individual measurements are shown by the horizontal grey marks, which include the contributions from both the NIRCam and the NIRSpec uncertainties. 
For a handful of galaxies, the NIRSpec continuum flux was low enough that the final measured flux did not have $S/N>3$, and we removed these this comparison.
From the large red points and errorbars, showing the median and standard deviation across all observations in a particular filter, we see that for all filters the median of the comparison points is well within one standard deviation of unity.
However, on average many of these indicate that the NIRCam fluxes are up to $\sim$10 \% lower than the NIRSpec fluxes.
In particular, we see evidence of a subtle trend with the NIRCam-to-NIRSpec flux ratio increasing slightly to long wavelengths. This arises as a result of the NIRSpec PSF degrading more significantly with increasing wavelength beyond 2.5 $\mu$m. 

Our NIRSpec data reduction pipeline does take this effect into account, in the form of the slitloss correction (which was undone for Figure~\ref{fig:flux_NS_NC_compare}). These slitlosses will depend on the morphology of the galaxy and its location within the shutter, but the $\sim$10 \% drift seen with wavelength in the comparison in Figure~\ref{fig:flux_NS_NC_compare} is entirely consistent with the differing PSFs of NIRCam and NIRSpec as a function of wavelength.
Slitlosses presented in this work were calculated assuming each galaxy is a point source at the location of the centroid derived from imaging.
Table~\ref{tab:sizes} gives the half-light radii from the JADES public catalog \citep{Rieke2023}\footnote{\url{https://archive.stsci.edu/hlsp/jades}} for the subset of our sample for which NIRCam imaging was available. In this study, we focused only on rest-frame optical emission lines at $z>5.5$, meaning that the observed-frame wavelengths were all $\lambda \gtrsim 2.5 \mu$m. At these wavelengths, the NIRCam PSF is typically worse than 0\farcs1 (0\farcs092 in F277W, 0\farcs116 in F356W\footnote{\url{https://jwst-docs.stsci.edu}}). Thus, at the wavelengths considered in this work, the assumption of a point source profile should be reasonable for all but a few of our largest targets.

The majority of the emission line ratios studied in this work have relatively short wavelength baselines, for which the precise slitloss corrections will make a negligible difference.
The $O32$ and $R23$ ratios are the notable exceptions to this. However, even before slitloss corrections are applied, Figure~\ref{fig:flux_NS_NC_compare} demonstrates that, on average, there is only a $\sim$10 \% effect across the full wavlength range. Given that \OIIll3726,3729 and \OIIIl5007 are only separated by $\sim$1.4 $\mu$m in the observed frame, the effect would be even smaller than this quoted 10 \%. Furthermore, the point-source slitloss corrections we do apply should be adequate for most sources. Thus, the final flux calibration uncertainty on the $O32$ ratio is likely well below 10 \% which is smaller than the quoted measurements uncertainties throughout this paper.

\begin{table}
\centering
\begin{tabular}{ccc}
\hline
ID & NIRCam ID & $R_{50}$ (\arcsec) \\
\hline
8013 & 110748 & 0.088 \\
4297 & 101683 & 0.090 \\
9422 & 113585 & 0.090 \\
20961 & 135819 & 0.096 \\
16625 & 127219 & 0.097 \\
10005113 & 110319 & 0.098 \\
21842 & 137667 & 0.099 \\
18846 & 131688 & 0.101 \\
19606 & 210963 & 0.111 \\
18179 & 130262 & 0.114 \\
6002 & 106292 & 0.114 \\
4404 & 101990 & 0.115 \\
6246 & 106885 & 0.116 \\
\hline
19342 & 132780 & 0.121 \\
3334 & 197088 & 0.126 \\
22251 & 138571 & 0.132 \\
18976 & 131971 & 0.136 \\
10013682 & 210625 & 0.143 \\
17566 & 209276 & 0.190 \\
16745 & 208642 & 0.191 \\
\hline
\end{tabular}
\caption{Half-light radii from the JADES public catalog \citep{Rieke2023} for the subset of our sample for which NIRCam imaging was available. `NIRCam ID' gives this cross-matched ID from those imaging catalogs. The horizontal rule marks the FWHM of the NIRCam PSF in the F356W filter (0\farcs116); i.e., targets below this rule have half-light radii larger than the NIRCam PSF.}
\label{tab:sizes}
\end{table}

\subsection{Comparison of grating flux calibration}

\begin{figure}
    \centering
    \includegraphics[width=0.97\columnwidth]{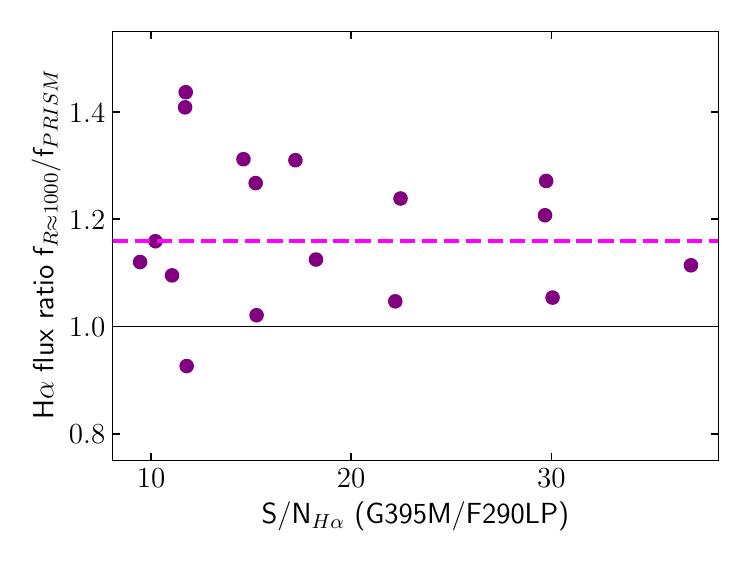}
    \caption{Comparison of H$\alpha$ flux measurements made from low-resolution PRISM/CLEAR and $R\sim1000$ G395M/F290LP measurements for galaxies in this sample. Fluxes reported in the $R\sim1000$ data are systematically higher by a median factor of 1.16.}
    \label{fig:flux_NS_NC_compare}
\end{figure}

This paper almost exclusively makes use of the PRISM/CLEAR mode.
This is because, at redshifts beyond $z>5$, the rest-frame optical emission lines fall beyond $\lambda_{\rm obs}\gtrsim2.5 \mu$m where the PRISM/CLEAR spectral resolution is $R\sim200-300$, meaning most of these lines are well resolved.
The notable exception to this is the [N {\sc ii}] $\lambda$6583 / H$\alpha$ ratio, for which we require the $R\sim1000$ G395M/F290LP grating.
Although reported ratios are only ever measured within the same spectral observation, we briefly consider the relative flux calibration of the PRISM/CLEAR and G395M/F290LP.
Figure~\ref{fig:flux_NS_NC_compare} shows the ratio of H$\alpha$ fluxes measured in the G395M/F290LP to those measured in the PRISM/CLEAR mode for galaxies in this sample which have spectral coverage in both settings. Note that the fluxes measured here are not those reported in the main text, where the continuum is modelled with stellar templates. Rather the continuum here is modelled with a simple spline fit. This is done to minimise any model-dependent effects in this comparison. The dashed magenta line shows the median value of 1.16, indicating that fluxes measured from the grating are systematically larger. This effect has been noted in \citet{Bunker2023_DR}. Note that in the PRISM/CLEAR, the reported H$\alpha$ flux would be blended with [N {\sc ii}], whereas these lines are resolved in the G395M/F290LP measurements. However, as pointed out in Section~\ref{sec:results}, [N {\sc ii}] $\lambda$6583 is not clearly detected in any of these galaxies, and this would contribute at the $<10$ \% level. Furthermore, the effect of this blending would be to increase the PRISM/CLEAR flux, but it is the G395M/F290LP flux that is determined to be larger.
As outlined above, the flux calibration in the PRISM/CLEAR mode is in good agreement with broadband photometry. Thus, we consider that the PRISM/CLEAR flux calibration is likely reliable, while the grating flux may be overestimated. However, given we only make use of the G395M/F290LP data to measure the ratio of two lines very close in wavelength within the same observation, the systematically high fluxes recorded should have no effect on any of the findings reported here.

\end{document}